\documentclass[preprint,superscriptaddress,amsmath,amssymb,longbibliography]{revtex4-1}  
\usepackage{graphicx}
\usepackage{bm}
\usepackage[colorlinks=true,urlcolor=blue,citecolor=blue,linkcolor=blue,bookmarks=false,pdfstartview={FitH}]{hyperref}      
\usepackage{color}
\usepackage{multirow}
\def \cm-1{cm$^{-1}$}

\begin{document}
\title{Magnetic quantum defects in a uniaxial antiferromagnetic insulator} 
\author{Shangfei~Wu}
\email{wusf@baqis.ac.cn}
\affiliation{Department of Physics and Astronomy, Rutgers University, Piscataway, New Jersey 08854, USA}
\affiliation{Beijing Academy of Quantum Information Sciences, Beijing 100193, China} 
\author{Laur~Peedu}
\affiliation{National Institute of Chemical Physics and Biophysics,
12618 Tallinn, Estonia}
\author{Zhihao Wang} 
\author{Xuecong Wang} 
\affiliation{Department of Physics and Astronomy and SmartState Center for Experimental Nanoscale Physics,
University of South Carolina, Columbia, South Carolina, 29208, USA}
\author{Xianghan~Xu}
\affiliation{Department of Physics and Astronomy, Rutgers University, Piscataway, New Jersey 08854, USA}
\affiliation{Keck Center for Quantum Magnetism, Rutgers University, Piscataway, New Jersey 08854, USA}
\author{Kai~Du}
\affiliation{Department of Physics and Astronomy, Rutgers University, Piscataway, New Jersey 08854, USA}
\affiliation{Keck Center for Quantum Magnetism, Rutgers University, Piscataway, New Jersey 08854, USA}
\author{Sang-Wook~Cheong} 
\affiliation{Department of Physics and Astronomy, Rutgers University, Piscataway, New Jersey 08854, USA}
\affiliation{Keck Center for Quantum Magnetism, Rutgers University, Piscataway, New Jersey 08854, USA}
\author{Aleksei~Boldin}
\affiliation{National Institute of Chemical Physics and Biophysics,
12618 Tallinn, Estonia}
\author{Joosep~Link}
\affiliation{National Institute of Chemical Physics and Biophysics,
12618 Tallinn, Estonia} 
\author{Ivo~Heinmaa}
\affiliation{National Institute of Chemical Physics and Biophysics,
12618 Tallinn, Estonia}
\author{Raivo~Stern}
\affiliation{National Institute of Chemical Physics and Biophysics,
12618 Tallinn, Estonia}
\author{Sai~Mu} 
\email{MUS@mailbox.sc.edu}
\affiliation{Department of Physics and Astronomy and SmartState Center for Experimental Nanoscale Physics,
University of South Carolina, Columbia, South Carolina, 29208, USA}
\author{Urmas~Nagel}
\affiliation{National Institute of Chemical Physics and Biophysics,
12618 Tallinn, Estonia}
\author{Toomas~Rõõm} 
\email{toomas.room@kbfi.ee}
\affiliation{National Institute of Chemical Physics and Biophysics,
12618 Tallinn, Estonia}
\author{Girsh~Blumberg} 
\email{girsh@physics.rutgers.edu}
\affiliation{Department of Physics and Astronomy, Rutgers University,
Piscataway, New Jersey 08854, USA}
\affiliation{National Institute of Chemical Physics and Biophysics,
12618 Tallinn, Estonia}
\date{\today}      

\maketitle

\[ \textbf{Abstract} \]

Point defects have been successfully utilized in various quantum technologies, serving as quantum qubits for quantum computation, single-photon emitters for quantum communication, and nanoscale sensors for quantum metrology. 
However, their further development faces key challenges, particularly in discovering and exploring suitable defect-host systems that meet the necessary criteria for quantum applications.
Here, using polarization-resolved Raman spectroscopy and terahertz absorption spectroscopy, we discover three distinct chromium-vacancy-induced \textcolor{black}{excitations} in the uniaxial antiferromagnetic insulator, Cr$_2$O$_3$. 
These \textcolor{black}{vacancy-induced excitations} have an energy scale of a few tens of millielectronvolts \textcolor{black}{and are twofold degenerate}, and the lowest one \textcolor{black}{at 64\,\cm-1} is sharp and sensitive to the external magnetic field along the easy-axis direction, particularly close to the spin-flop regime around 6\,T, where the mode softens from 64 to 27\,\cm-1.
\textcolor{black}{Based on the defect supercell first-principles calculations, we interpret the mode at 64\,\cm-1 as a local magnetic excitation of the local moment within the chromium vacancy state.}
Our results establish that the magnetic defect states in Cr$_2$O$_3$ \textcolor{black}{have potential for quantum applications.}

\newpage

\textbf{Introduction}

Point defects are localized at or near a single lattice site and are classified as zero-dimensional defects. They are commonly seen as 
vacancies, interstitials, antisites, substitutional atomic impurities, and complexes like a vacancy encountering an impurity, in all materials~\cite{COWLEY_1972_RevModPhys,crawford1975point,Tuller_Point_Defects_in_Oxides2011,mccluskey2012dopants,Zhang2012,Freysoldt_2014_RevModPhys,Dreyer_2018_annurev,Dreyer_2024_JAP}. While defects are often detrimental to device performance, in some cases, a small percentage of defect concentrations can significantly change material and device properties. 
For example, introducing dopants into silicon can drastically affect the electrical properties~\cite{KOHN_1957,Bassani_1974}, forming the base of the modern semiconductor industry.
More recently, defects have emerged as robust and manipulable quantum systems for the next generation of quantum technologies, e.g., spin qubits for quantum computation~\cite{Kane1998,Pla2012,Wu2019}, single-photon emitters for quantum communication~\cite{aharonovich_2011_diamond,Aharonovich_2011_Nature_Photonics}, and nanoscale sensors for quantum metrology~\cite{Schirhagl_2014_review}. 
To further advance the development of defect-based quantum technologies, it is essential to explore suitable defect-host systems that meet the necessary criteria for quantum applications~\cite{Weber_2010_PNAS,Childress_2014_Physics_Today,Nathalie_2021_Science,Degen_2017_RevModPhys}.   
                       
In this work, we present evidence for the discovery of three distinct chromium vacancy-induced \textcolor{black}{excitations} in the uniaxial antiferromagnetic (AFM) insulator, Cr$_2$O$_3$. The lowest-energy defect-induced excitation \textcolor{black}{at 64\,\cm-1} is \textcolor{black}{sharp} and sensitive to the external magnetic field along the easy-axis direction close to the spin-flop regime around 6\,T, where it softens from 64 to 27\,\cm-1. \textcolor{black}{The mode at 64\,\cm-1 is interpreted as a local magnetic excitation of the local moment within a chromium vacancy state based on our defect supercell first-principles calculations.}
These results motivate the use of chromium vacancies in Cr$_2$O$_3$ as \textcolor{black}{magnetic quantum defects for quantum applications.}

\begin{figure*}[!t] 
\begin{center}
\includegraphics[width=\columnwidth]{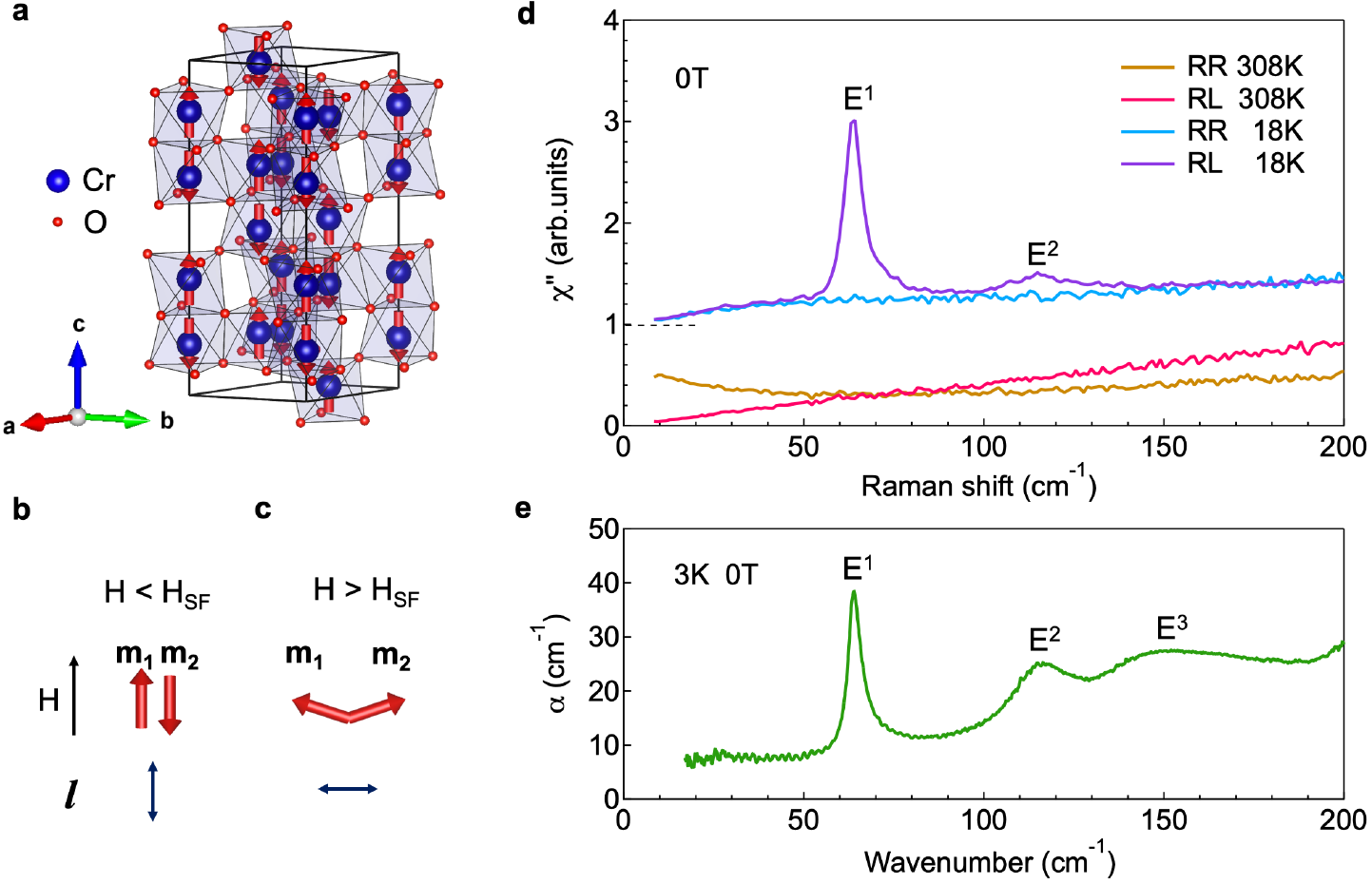}
\end{center}
\caption{\label{Fig1} 
\textbf{Low-energy Raman and absorption spectra for the ``Rutgers"-Cr$_2$O$_3$ sample.} 
\textbf{a} Illustration of Cr$_2$O$_3$ spin structure in a hexagonal unit cell. 
\textbf{b} Illustration of the collinear AFM spin structure for magnetic field along the $c$ axis below the spin-flop field $H<H_{\text{SF}}$. \textbf{c} Same as panel \textbf{b} but for $H>H_{\text{SF}}$. $H$ is the external magnetic field along the $c$ axis.
$\bold{m}_1$ and $\bold{m}_2$ denote the sublattice magnetization unit vector. $\textit{\textbf{l}}=\bold{m}_1-\bold{m}_2$ denotes the staggered magnetization vector (or vector of the AFM order). It is two-fold degenerate with respect to the sign of the staggered vector $\textit{\textbf{l}}$, which is represented by two-headed arrows in the bottom panel~\cite{Bogdanov_2007_PhysRevB}.
\textbf{d} Low-energy Raman response for the ``Rutgers"-Cr$_2$O$_3$ at 308\,K and 18\,K from the $ab$ plane in both $RR$ [$A_1+A_2$] and $RL$ [$E$] scattering geometries at 0\,T.
\textbf{e} Low-energy terahertz absorption spectra from the $ab$ plane of ``Rutgers"-Cr$_2$O$_3$ at 3\,K and 0\,T. The terahertz field propagates along the $c$ axis, while the electric and magnetic field components of the terahertz wave are in the $ab$ plane.}
\end{figure*}

\begin{figure*}[!ht] 
\begin{center}
\includegraphics[width=1\columnwidth]{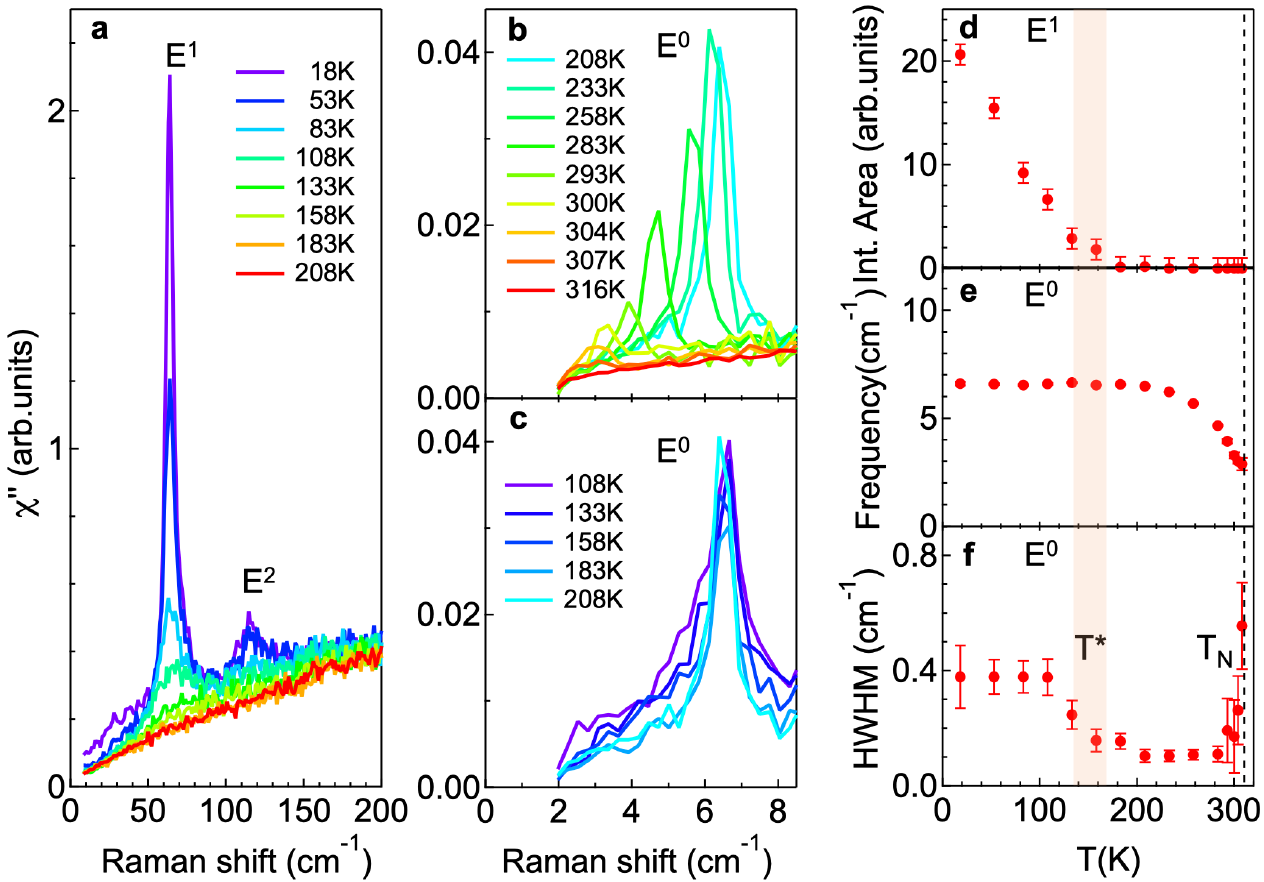}
\end{center}
\caption{\label{Fig2} 
\textbf{Temperature dependence of the Raman response for the ``Rutgers"-Cr$_2$O$_3$ sample in the $RL$ scattering geometry at 0\,T.}
\textbf{a} Temperature dependence of the Raman response for the $E^1$ and $E^2$ modes.  
\textbf{b-c} Temperature dependence of the Raman response for the one-magnon scattering peak $E^0$ below 8\,\cm-1.
\textbf{d} Temperature dependence of the integrated area for the $E^1$ mode. 
\textbf{e} Temperature dependence of the mode frequency for the one-magnon scattering peak $E^0$. 
\textbf{f} Same as panel \textbf{e} but for the half-width at half-maximum (HWHM). The orange shaded area in panels \textbf{d-f} denotes the temperature $T^*\sim 150$\,K, while the vertical dashed line in panels \textbf{d-f} denotes the temperature $T_\text{N}$.}
\end{figure*}   
    
\textbf{Results}
   
\textbf{\textcolor{black}{Low-energy modes}.}

Cr$_2$O$_3$ is the first discovered material that shows a linear magnetoelectric effect~\cite{Fiebig_2005_review}. It is an band insulator with a band gap of 3.4\,eV~\cite{adler1968insulating,crawford1964electricalCr2O3,zaanen1985bandgap}. It belongs to a corundum (hexagonal) crystal structure [Fig.~\ref{Fig1}a] with space group $R\bar{3}c$ (No.~167) (point group $D_{3d}$) above $T_\text{N}=307.5$\,K (Supplementary Figure~\ref{Fig.S1}).
\textcolor{black}{
According to the group-theoretical analysis of the bulk hexagonal Cr$_2$O$_3$, Raman-active phonon modes at the Brillouin zone center are $\Gamma_{\text{Raman}}$= 2$A_{1g}$ $\oplus$ 5$E_g$ (Supplementary Note~\ref{GroupTheory}). These modes 
are detected in our Raman experiment that two $A_{1g}$ modes are at around 295 and 552\,\cm-1, and that five $E_g$ modes are at around 292, 350, 394, 528, 615\,\cm-1 at 308\,K~(Supplementary Figure~\ref{Fig.S2}), consistent with a previous Raman study~\cite{Shim_2004_PhysRevB}.}

Below $T_\text{N}$, Cr$_2$O$_3$ forms a two-sublattice collinear AFM order (magnetic space group $R\bar{3}'c'$) with easy axis along the $c$ axis as shown in Figs.~\ref{Fig1}a-b~\cite{Brockhouse_1953,Corliss_1965}. 
When an external magnetic field is applied along the $c$ axis, a spin-flop transition occurs at approximately 6\,T~\cite{Foner_1963_PhysRev}. Above this critical field, the Cr spins are canted in the $ab$ plane with a net magnetization along the $c$ axis shown in Fig.~\ref{Fig1}c~\cite{Fiebig_1996_PhysRevB,Popov_1999,Li_2020_nature}.
The presence of AFM ordering in the bulk Cr$_2$O$_3$ breaks inversion ($i$), dihedral mirror ($\sigma_d$), and rotoinversion ($S_6$) symmetries associated with the $D_{3d}$ point group, reducing the system to the $D_3$ point group~\cite{Ren_2024_PhysRevX} (Supplementary Note~\ref{GroupTheory}).
In the following, we use the irreducible representations of the $D_3$ point group to discuss the low-energy excitations below $T_\text{N}$.
      
In Fig.~\ref{Fig1}d, we show the low-energy Raman response below 200\,\cm-1 recorded from the $ab$ plane of the ``Rutgers"-Cr$_2$O$_3$ sample \textcolor{black}{(details in Methods)}. 
At 308\,K, we detect quasielastic scattering in the $RR$ scattering geometry, suggesting the existence of the fully-symmetric magnetic fluctuations above $T_\text{N}$ (Supplementary Note~\ref{QEP}). Below $T_\text{N}$, the quasielastic scattering goes away, as shown in the 18\,K data.
In the $RL$ scattering geometry, we detect a flat continuum response in the 308\,K data. At 18\,K, two modes are detected. One is located at 64\,\cm-1 (labeled as $E^1$) and the other one is located at 115\,\cm-1 (labeled as $E^2$). 
We also confirm these two modes at the same peak positions in the terahertz absorption spectra at 3\,K shown in Fig.~\ref{Fig1}e. Additionally, a third peak located at about 150\,\cm-1 (labeled as $E^3$) is observed in the terahertz absorption spectra, but not in the Raman spectra. This might be because the $E^3$ mode is too weak to be detected in the Raman experiment. 
\textcolor{black}{
Since these three modes ($E^1$, $E^2$, and $E^3$) are well separated from the bulk phonon modes~(Supplementary Figure~\ref{Fig.S2}), we regard them as three additional modes.}
              
\textbf{Temperature dependence.}   

To figure out at what temperature these additional modes appear, we show the temperature dependence of the Raman response for the $E^1$ and $E^2$ modes in the $RL$ scattering geometry at 0\,T in Fig.~\ref{Fig2}a. These two modes gain spectral weight below around 150\,K, and get stronger and narrower upon further cooling.  The temperature dependence of the integrated area for the $E^1$ mode is shown in Fig.~\ref{Fig2}d. These results suggest an onset temperature of $T^*\sim150$\,K for these additional modes to emerge. 

In Figs.~\ref{Fig2}b-c, we show the Raman response below 8\,\cm-1 in the $RL$ scattering geometry at 0\,T. Once the sample is cooled just below $T_\text{N}$, a broad and weak mode appears at 2.8\,\cm-1 (labeled as $E^0$). It shifts to higher energies upon further cooling, saturates to 6.5\,\cm-1 at 200\,K, and its peak position barely changes upon further cooling to 18\,K. This mode is interpreted as the one-magnon scattering peak at the Brillouin zone center, consistent with previous inelastic neutron and Raman scattering studies~\cite{SAMUELSEN_1968,Samuelsen_1969_Solid,DongBiao_2023_PIP}. 
Furthermore, the mode gets sharper and shows an asymmetric line shape upon cooling below $T_\text{N}$ [Fig.~\ref{Fig2}b]. To our surprise, as we show in Fig.~\ref{Fig2}c, this magnon mode gets broadened below around 150\,K, roughly the same temperature $T^*$ when the $E^1$ mode appears. We use a Fano model to capture the coupling between the one-magnon mode and a continuum at low energy in the same symmetry channel (Supplementary Note~\ref{fitting_magnon}).
The temperature dependence for the one-magnon mode's frequency and HWHM are shown in Figs.~\ref{Fig2}e and \ref{Fig2}f, respectively.
The concurrence of the broadening of the one-magnon peak and the appearance of the additional modes below $T^*$ suggests that an additional interaction takes effect on the Cr sites and affects Cr spins at a long distance. Since the threefold rotational symmetry is preserved for these additional modes in the AFM phase (Supplementary Note~\ref{three_fold_rotational_symmetry}), we hypothesize that the additional interaction is related to Cr vacancies, because Cr vacancies reside on the high symmetry threefold axis.

\textbf{Cr vacancy dependence.}  

\begin{figure}[p] 
\begin{center}
\includegraphics[width=\columnwidth]{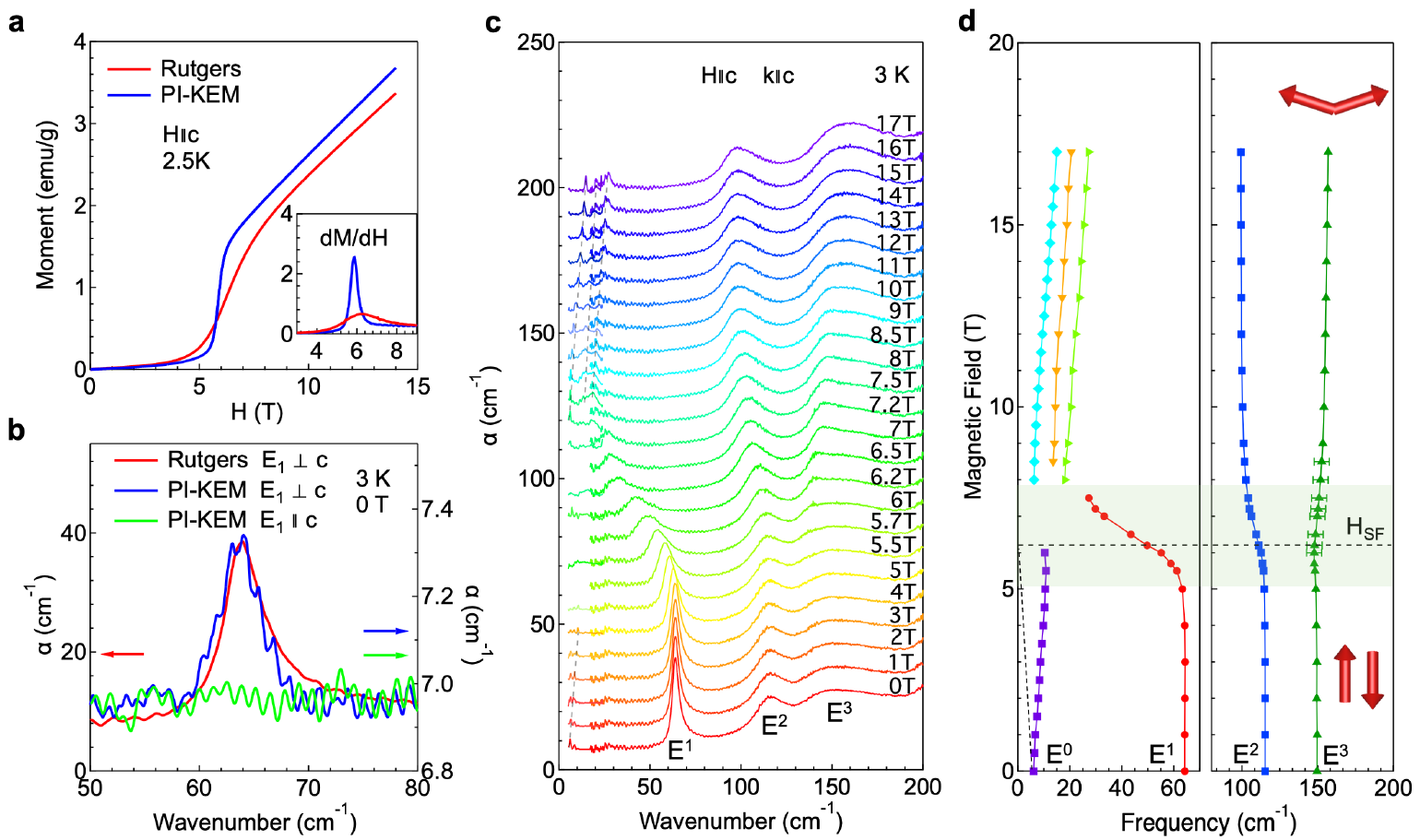}
\end{center}
\caption{\label{Fig3} 
\textbf{Magnetic field dependence.}
\textbf{a} Magnetic field dependence of the magnetization data for ``Rutgers" and ``PI-KEM" Cr$_2$O$_3$ samples at 2.5\,K and $H||c$.  The inset of panel \textbf{a} shows the derivative of magnetization with respect to the magnetic field $dM/dH$.
\textbf{b} $E^1$ mode in the absorption spectra for the ``Rutgers" and ``PI-KEM" Cr$_2$O$_3$ sample at 3\,K and 0\,T. 
The left axis shows the terahertz absorption spectra from the $ab$ plane of the “Rutgers” Cr$_2$O$_3$ sample at 3\,K, where the terahertz field propagates along the $c$-axis direction. The electric field $E_1$ and magnetic field $H_1$ components of the terahertz wave are in the $ab$ plane. The right axis shows the data for the ``PI-KEM" Cr$_2$O$_3$ sample from the $bc$ plane, where the components of the terahertz wave are $E_1 \perp c$ and $H_1 \parallel c$ (blue curve);
$E_1 \parallel c$ and $H_1 \perp c$ (green curve).
\textbf{c} Magnetic field dependence of the terahertz absorption spectra from the $ab$ plane of the “Rutgers”-Cr$_2$O$_3$ sample at 3\,K in the Faraday configuration in two frequency ranges, 5-23\,\cm-1 and 18-200\,\cm-1. The dashed lines are guides to the eyes. The complete set of the same absorption spectra in the 5-23\,\cm-1 range is shown in Supplementary Figure~\ref{Fig.S7}. 
The data are vertically shifted for clarity.
\textbf{d} Magnetic field dependence of the mode frequency for the $E^0$, $E^1$, $E^2$, and $E^3$ modes. The horizontal dashed line represents the spin-flop field $H_\text{SF}$. The shaded area denotes the spin-flop field regime.
The dashed line below 10\,\cm-1 represents the expected lower branch of the  magnon mode $E^0$.
}
\end{figure}

To validate this hypothesis, \textcolor{black}{we compare the magnetization data and the terahertz absorption spectra for the two Cr$_2$O$_3$ samples: the ``Rutgers" and ``PI-KEM" samples (details in Methods).}
In Fig.~\ref{Fig3}a, we show the temperature dependence of the magnetization data  for the ``Rutgers" and ``PI-KEM" Cr$_2$O$_3$ samples in an applied magnetic field along the $c$ axis. The ``PI-KEM" sample has a sharper spin-flop transition. This is also seen from the $dM/dH$ data shown in the inset of Fig.~\ref{Fig3}a.
\textcolor{black}{Since the Cr atoms are the magnetic ions, the width of the spin-flop transition reflects the sample inhomogeneity, which is proportional to the amount of the Cr vacancies.} 
Thus, the sharper spin-flop transition indicates fewer Cr vacancies for the ``PI-KEM" sample compared with the ``Rutgers"-Cr$_2$O$_3$ sample. 
The fewer Cr vacancies in the ``PI-KEM" Cr$_2$O$_3$ sample are also supported by the absence of the upturn below 20\,K in the magnetic susceptibility measurements [Supplementary Figure~\ref{Fig.S1}b].
Furthermore, we compare the $E^1$ mode strength in the terahertz absorption spectra for these two samples. 
The $E^1$ mode appears only when the electric field component $E_1$ of the terahertz wave is in the $ab$ plane. As shown in Fig.~\ref{Fig3}b, the $E^1$ mode strength is about $\sim$100 times weaker in the ``PI-KEM"  sample compared with the ``Rutgers" sample. \textcolor{black}{Since the impurity mode intensity is directly related to the impurity \textcolor{black}{concentration}}, the appearance of the additional modes is related to the existence of Cr vacancies in Cr$_2$O$_3$, and the strength of these additional modes depends on the amount of Cr vacancies.

\textbf{Magnetic field dependence.}  

To further understand the origin of these additional modes, we present the magnetic field dependence of the three modes ($E^1$, $E^2$, and $E^3$) at 3\,K for the ``Rutgers"-Cr$_2$O$_3$ sample in the Faraday configuration, shown in Fig.~\ref{Fig3}c.
When the magnetic field is below the spin-flop field $H_\text{SF}$=6.2\,T~\cite{Dayhoff_1957_PhysRev,Foner_1963_PhysRev}, the three modes ($E^1$, $E^2$, and $E^3$) barely change, and no splittings are detected. 
Close to the spin-flop transition in the field range between 5 and 7\,T, the three additional modes change drastically. The $E^1$ mode softens, broadens, and weakens upon approaching $H_\text{SF}$, and finally disappears above $H_\text{SF}$. $E^2$ and $E^3$ modes repel each other, and their energy difference becomes larger and saturates at higher magnetic fields. 
The high-energy branch of the magnon mode $E^0$ shifts linearly in the magnetic field below $H_\text{SF}$ and disappears close to $H_\text{SF}$. 
Above the spin-flop field $H_\text{SF}$, the quasi-ferromagnetic mode emerges and shifts to higher energies with increasing magnetic field, reaching 15\,\cm-1 at 17\,T.
Two more modes appear above $H_\text{SF}$, shift to higher energies with increasing magnetic field, and reach 20.5\,\cm-1 and 27.2\,\cm-1at 17\,T.
The detailed magnetic field dependence of these modes' frequencies for the $E^0$, $E^1$, $E^2$, and $E^3$ modes is presented in Fig.~\ref{Fig3}d.

                                                                                                                                                                                                                                                                                                                                                                                                                                                                                                                                                                                                                                                                                                                
\textbf{Discussions}
     
We discuss the origin of the three additional modes ($E^1$, $E^2$, and $E^3$). They are not phonon modes \textcolor{black}{for a pristine crystal} because the lowest optical phonon mode at the Brillouin zone center for Cr$_2$O$_3$ is at about 33.3\,meV (266\,\cm-1) based on the DFT phonon calculation~\cite{Ren_2024_PhysRevX}. 
They are not the one-magnon modes, because the one-magnon modes at the Brillouin zone center are around 6.5\,\cm-1 and 400\,\cm-1~\cite{DongBiao_2023_PIP,SAMUELSEN_1968,Samuelsen_1969_Solid}. 
\textcolor{black}{They are not consistent with the magnon-polaron modes suggested by Ref.~\cite{Li_2020_PhysRevLett}, 
because the three additional modes already exist at 0\,T.
The scenario of the bound magnetic polaron mode~\cite{Franchini_2021_review,alexandrov2010advances} is not consistent with our data, because
Cr$_2$O$_3$ is a band insulator with a band gap of 3.4\,eV~\cite{adler1968insulating,crawford1964electricalCr2O3,zaanen1985bandgap}, a low-density charge carrier is absent to be coupled with the localized spins of magnetic ions to form this type of mode. Furthermore, the peak position of a magnetic polaron mode shifts in the magnetic field~\cite{Rho_2002_PhysRevLett}, which is inconsistent with our data below $H_\text{SF}$ [Figs.~\ref{Fig3}c,d].}
    
\begin{figure}[t] 
\begin{center}
\includegraphics[width=0.6\columnwidth]{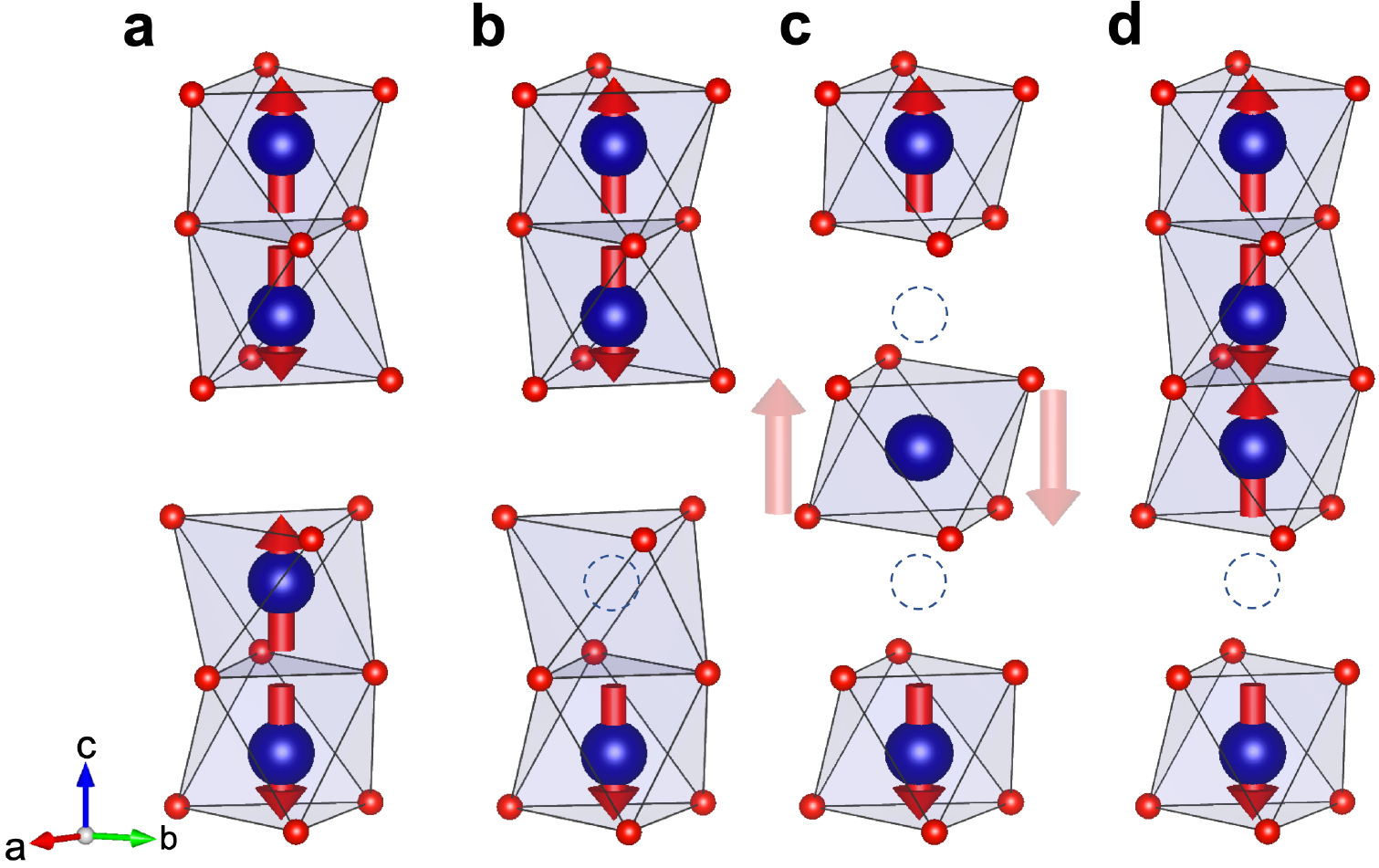}
\end{center}
\caption{\label{Fig4} 
\textbf{Illustrations of different types of Cr vacancies.} 
\textbf{a} The Cr$_2$O$_3$ crystal structure without Cr vacancies. 
\textbf{b} Conventional Cr vacancy $V_\text{Cr}$ that one Cr atom is absent at the Cr site. 
\textbf{c} Split Cr vacancies $V^{(s)}_\text{Cr}$ where a nearby Cr undergoes a significant relaxation in a conventional Cr vacancy shown in panel \textbf{b}, migrating along the hexagonal axis to an interstitial site (the inversion center) in between, thereby forming a dumbbell structure along the hexagonal axis. It is equivalent to the formation of an interstitial Cr together with two Cr vacancies. The spin direction for the interstitial Cr atom could be either parallel or antiparallel to the $c$ axis.
\textbf{d} Defect complex $\text{Cr}_{i} \text{--} V_\text{Cr}$, where a Cr atom is displaced from its lattice site and relocates to a nearby interstitial position.}
\end{figure}

One way to interpret the additional modes is that they represent localized lattice vibration modes resulting from the presence of Cr vacancies.  
Since the “Rutgers”-Cr$_2$O$_3$ samples are grown involving air flow~\cite{Du_2023_NPJ} and the “PI-KEM” Cr$_2$O$_3$ samples are grown involving O$_2$ flow, the synthesis environments for both samples are under O-rich conditions, where Cr vacancies are expected to be present in the sample.
According to first-principles calculations~\cite{Sai_Mu_2025_arxiv}, Cr-vacancy states are deep acceptor states. They generally have lower formation energies
under O-rich conditions than under Cr-rich conditions. 
Three types of Cr-vacancies preserve the threefold rotational symmetry. The first one is a conventional Cr vacancy state $V_\text{Cr}$, where one Cr atom is absent at the Cr site [Fig.~\ref{Fig4}b]. 
The second one is the split Cr vacancy $V^{(s)}_\text{Cr}$, which requires a nearby Cr migrating along the hexagonal axis to an interstitial site (the inversion center) in a conventional Cr vacancy state $V_\text{Cr}$, equivalent to the formation of an interstitial Cr together with two Cr vacancies~[Fig.~\ref{Fig4}c].
The third one is the defect complex $\text{Cr}_{i} \text{--} V_\text{Cr}$, where a Cr atom is displaced from its lattice site and relocates to a nearby interstitial position (the inversion center)~[Fig.~\ref{Fig4}d].
Compared with the conventional Cr vacancy $V_\text{Cr}$ and the defect complex $\text{Cr}_{i} \text{--} V_\text{Cr}$, the split Cr vacancy $V^{(s)}_\text{Cr}$ has the lowest formation energy, thus is most likely to form in the bulk sample~\cite{Sai_Mu_2025_arxiv}. 
Our supercell DFT phonon calculations based on the split Cr vacancy $V^{(s)}_\text{Cr}$ find local $E$-symmetry vibration modes at 130\,\cm-1 and 150-200\,\cm-1, close to the energy of the $E^2$ and $E^3$ modes, suggesting that $E^2$ and $E^3$ modes are local vibration modes~(Supplementary Figure~\ref{Fig.S8}). The small changes for the $E^2$ and $E^3$ modes across $H_\text{SF}$ could be due to a tiny field-induced change of the $c$-axis lattice constant~\cite{YACOVITCH19771126}. 
\textcolor{black}{
For the $E^1$ mode, the large softening near $H_\text{SF}$~(Fig.~\ref{Fig3}d) and the absence of any local vibration modes below 100\,\cm-1 from the DFT defect supercell calculation~(Supplementary Figure~\ref{Fig.S8}) suggest that it is not likely to be a local lattice vibration mode.
}

An alternatative way to interpret the $E^1$ mode involves a local magnetic excitation associated with the split Cr vacancy $V^{(s)}_\text{Cr}$.
In insulating Cr$_2$O$_3$ samples, $V^{(s)}_\text{Cr}$ is in the $-$3 charge state and carries a local moment of 2.7\,$\mu_\text{B}$ on the interstitial Cr, which is located \textit{near} the inversion center~\cite{Sai_Mu_2025_arxiv}. 
While the effective exchange field vanishes exactly at the inversion center in the AFM phase of Cr$_2$O$_3$, the exchange field on the interstitial Cr \textit{near} the inversion center is small, indicating that the orientation of its local moment can flip with little resistance.
\textcolor{black}{
Indeed, the energy cost to flip the spin on the interstitial Cr is only about 15\,meV~\cite{Sai_Mu_2025_arxiv}.  
In the mean field limit, this energy cost for a spin flip for $S=3/2$ corresponds to a magnetic excitation between $S_z=+3/2$ and $S_z=-3/2$ states, with spin quantum number change of $\Delta S_z=3$. 
For a local magnetic excitation with spin quantum number change of $\Delta S_z=1$ induced by light, the excitation energy is expected to be lower than 15\,meV. 
We further build up a spin Hamiltonian for the spin at the interstitial Cr --- with parameters calculated from DFT, incorporate the elastic interactions, and estimate the excitation energy in the mean field limit (Supplementary Note~\ref{local_spin_flip_enegy}).
We found that the energy difference between the minimum of the $S_z = \pm 3/2$ branch and the minimum of the $S_z = \pm 1/2$ branch is 11 meV (Supplementary Note~\ref{local_spin_flip_enegy}). 
This corresponds to a local magnetic excitation with spin quantum number change of $\Delta S_z=1$ induced by the terahertz light.
While this energy is about 1.4  times the $E^1$ mode we observed in the experiment, we notice that the $E^1$ mode shows the asymmetric lineshape both in the Raman and absorption spectra (Figs.~\ref{Fig1}d-e), suggesting the existence of coupling between the interstitial Cr spin and the surrounding O atoms.} 
\textcolor{black}{Furthermore, the spin-up state for the interstitial Cr could tunnel to the spin-down state near the inversion center and fluctuate rapidly, as a result of a few millielectronvolt potential barrier and the tiny deviations from the inversion center. This explains the selection rule and weak field depdendence below $H_\text{SF}$ for the $E^1$ mode (Supplementary Note~\ref{tunneling}).}
The large softening of the $E^1$ mode across $H_\text{SF}$ can be due to the change of the local effective magnetic field at the inversion center, driven by the external magnetic field along the easy-axis $c$ direction. 
\textcolor{black}{For zero external magnetic field or low magnetic field below 5\,T, the magnetic order in Cr$_2$O$_3$ remains collinear AFM order (Fig.~\ref{Fig1}b). The effective magnetic field at the inversion center is zero due to the perfect sublattice cancellation.}
In the spin-flop regime, the effective magnetic field at the inversion center is finite due to the reorientation of the spins and quasi-ferromagnetic nature along the $c$ axis~(Fig.~\ref{Fig1}c).
For Cr$_2$O$_3$, the rotation of the staggered magnetization vector at the regular Cr site is clear only in the spin-flop regime (5-7\,T) as shown in Fig.~\ref{Fig3}a. 
Consequently, the resulting change of the effective magnetic field at the inversion center is evident in the same external magnetic field range. This change of the local magnetic field environment for the interstitial Cr atom could drive the softening of the $E^1$ mode.

                              
\textbf{Summary} 
              
In summary, we identify three distinct twofold-degenerate chromium-defect-induced excitations in Cr$_2$O$_3$ by polarization-resolved Raman and terahertz absorption spectroscopies. The energy of these defect-induced excitations, especially the $E^1$ mode at 64\,\cm-1, is sensitive to the external $c$-axis magnetic field in the spin-flop regime around 6\,T, where it softens from 64 to 27\,\cm-1. 
\textcolor{black}{According to the defect supercell first-principles calculations, the mode at 64\,\cm-1 is interpreted as a local magnetic excitation of the moment within the chromium vacancy state.}
\textcolor{black}{Our results establish that chromium vacancies in Cr$_2$O$_3$ can be used as magnetic defects for quantum applications~\cite{Schirhagl_2014_review,Degen_2017_RevModPhys}.}
     
\textbf{Methods}

\textbf{Single crystal preparation and characterization.}

Single crystals of Cr$_2$O$_3$ were synthesized by a floating zone technique described in Ref.~\cite{Du_2023_NPJ} at Rutgers University. 
They were labeled as ``Rutgers"-Cr$_2$O$_3$. 
The (001) plane of the ``Rutgers"-Cr$_2$O$_3$ was polished with a lapping film (1\,$\mu$m, Buehler).
Another single crystal of Cr$_2$O$_3$ was purchased as polished with (100) plane from PI-KEM Ltd. This sample was grown by Verneuil method and labeled as ``PI-KEM" Cr$_2$O$_3$.
Both samples were characterized by magnetic susceptibility measurements with a magnetic field $H=100$\,Oe along the $c$ axis~[Supplementary Figure~\ref{Fig.S1}]. They have similar AFM phase transition temperatures, which are extracted to be close to $T_\text{N}=307.5$\,K~[inset of Supplementary Figure~\ref{Fig.S1}a]. The slight upturn in the magnetic susceptibility data below 20\,K for the ``Rutgers"-Cr$_2$O$_3$ sample, which is absent in the ``PI-KEM"-Cr$_2$O$_3$ sample [Supplementary Figure~\ref{Fig.S1}b], suggests more defects in the ``Rutgers"-Cr$_2$O$_3$ sample.
The sharp Raman modes and the low residual spectra background [Supplementary Figure~\ref{Fig.S2}] indicates the high quality of the single crystals.

\textbf{Raman scattering measurements.}\label{Raman}        
                                   
The ``Rutgers"-Cr$_2$O$_3$ samples were polished in the air to expose a (001) crystallographic plane.  
A strain-free area was examined by a Nomarski image. The polished crystals were positioned in
a continuous helium flow optical cryostat. The Raman measurements
were mainly performed using the Kr$^+$ laser line at 647.1\,nm (1.92\,eV) in
a quasibackscattering geometry along the crystallographic $c$ axis.
The excitation laser beam was focused into a $50\times100$ $\mu$m$^2$
spot on the $ab$ surface, with the incident power around 10\,mW. The
scattered light was collected and analyzed by a triple-stage Raman
spectrometer and recorded using a liquid-nitrogen-cooled
charge-coupled detector. 
Linear and circular polarizations were used in this study to decompose the Raman data into different irreducible representations.
The instrumental resolution was maintained better than 1.5\,\cm-1 for the measurement range from 8 to 700\,\cm-1, while it was maintained better than 0.5\,\cm-1 for the one-magnon mode measurement below 8\,\cm-1. 
All linewidth data presented were corrected for the instrumental resolution. 
The temperatures were corrected for laser heating (Supplementary Note~\ref{laser_heating_determination}). 
                                                                                                                                                                                                                                                                                                                                                                                                                                                                           
All spectra shown were corrected for the spectral response of the spectrometer and charge-coupled detector to obtain the Raman intensity $I
_{\mu v}$, which is related to the Raman response $\chi''(\omega,T)$: $I_{\mu v}(\omega, T)=[1+n(\omega, T)] \chi_{\mu \nu}^{\prime \prime}(\omega, T)$. Here $\mu (v)$ denotes the polarization of the 
incident (scattered) photon, $\omega$ is energy, $T$ is temperature, and $ n(\omega, T)$ is the Bose factor.
                                                                                                                                                                                           
The Raman spectra were recorded from the $ab$ (001) surface for scattering geometries denoted as $\mu v = XX, XY, RR, RL$, which is short for $Z(\mu v)\bar{Z}$ in Porto’s notation, where $X$ and $Y$ denote linear polarization parallel and perpendicular to the crystallographic axis, respectively; $R=X+iY$ and $L=X-iY$ denote the right- and left-circular polarizations, respectively. The $Z$ direction corresponds to the $c$ axis perpendicular to the (001) plane. 
                        
\textbf{Terahertz absorption measurements.}\label{IR}

The ``Rutgers"-Cr$_2$O$_3$ sample for the optical absorption experiments has $ab$ surface of $3.5 \times 3.5$ mm$^2$ and a thickness of 0.85\,mm. The ``PI-KEM"-Cr$_2$O$_3$ sample for the optical absorption experiments has $bc$ surface of $5 \times 3$ mm$^2$ and a thickness of 1.1\,mm. 
The absorption spectra were recorded using a Sciencetech SPS200 Martin-Puplett type spectrometer with a 0.3 K bolometer and a rotatable polarizer in front of the sample. Sample chamber was inside the cold bore of the superconducting 17\,T solenoid. 
        
\textbf{Magnetic susceptibility measurements.}\label{Magnetic_susceptibility} 

Magnetic susceptibility was measured using the vibrating-sample magnetometer (VSM) of the Quantum Design 14T-PPMS in magnetic fields ranging from 100\,Oe to 14\,T and temperatures down to 1.8\,K. At low magnetic fields, the temperature  dependence of magnetic susceptibility was measured in both the zero-field-cooled (ZFC) and field-cooled (FC) regimes. Magnetic field was applied along the easy axis.

\textbf{Density functional theory calculations.}\label{Raman}                                                                                                       
\label{DFT_phonon_calculaton}  
Density functional theory (DFT) calculations were performed using the Vienna Ab initio Simulation Package ~\cite{Kresse1996_PhysRevB,KRESSE199615}. The calculation details are in consistent with Wang {\it{et al.}}~\cite{Sai_Mu_2025_arxiv}. Projector augmented wave method~\cite{blochl1994projector} was employed and the exchange-correlation was treated within generalized gradient approximation (GGA)~\cite{Perdew1996PhysRevLett}. 
A cutoff energy for plane wave expansion of 520\,eV was used. A 10-atom rhombohedral primitive cell of Cr$_2$O$_3$ was adopted and fully relaxed using a 3$\times$3$\times$3 $\mathit{\Gamma}$-centered$~\mathit{k}$-point mesh. 
To account for the Coulomb correlation of the localized $d$ orbitals on Cr, we employed the hybrid functional of Heyd, Scuseria, and Ernzerhof (HSE)~\cite{heyd2003hybrid} with a mixing parameter $\alpha = 0.20$ and a screening parameter of $0.3$. The combination of these parameters yields a band gap of 3.45\,eV, in good agreement with the experimental value (3.4\,eV)~\cite{adler1968insulating,crawford1964electricalCr2O3,zaanen1985bandgap}.   

Defects were simulated using the supercell approach~\cite{freysoldt2014first,van2004first}. A 2~$\times$~2~$\times~$1 hexagonal supercell (120 atoms) of Cr$_2$O$_3$ was constructed and a single defect was introduced therein. Atomic relaxation was performed using a $\Gamma$ point sampling until the Hellemann-Feynman force on all atoms were reduced below 0.01\,eV/\AA.  

Interatomic force constants of the defective supercells are calculated using the finite displacement method as implemented in the \uppercase{phonopy} code~\cite{TOGO20151}. To check the lattice vibrations arisen from the defect, we unfold the phonon bands from the Brillouin zone of hexgonal supercell to rhombohedral primitive cell using the unfolding method~\cite{ikeda2017mode}. 
              
\textbf{Group-theoretical analysis.}

Group theoretical analysis was performed using the tools provided in the Bilbao Crystallographic Server~\cite{Bilbao_1, Bilbao_4}. 
The information about the irreducible representations of point groups and space groups follow the notations of Cracknell~$et~al$~\cite{cracknell1979general}.

\textbf{Data availability}\\ 
Data in this study are available from the corresponding authors upon request.

\bibliographystyle{naturemag}

 \newpage   
\vspace{1cm}
\textbf{Acknowledgments} \\
We thank Sandor Borda{\'c} for making the PI-KEM sample available.
The \textcolor{black}{Raman} spectroscopic work conducted at Rutgers (S.W. and G.B.) was supported by NSF Grant No.~DMR-2105001. 
\textcolor{black}{The THz work at NICPB (L.P., U.N., T.R., and G.B.) was supported by the European Research Council (ERC) under the European Union’s Horizon 2020 research and innovation programme grant agreement No.~885413.}
The work at BAQIS (S.W.) was supported by the National Natural Science Foundation of China (Grant No.~12404548).
The sample growth, characterization, and polishment work (X.X., K.D, and S.-W.C.) was supported by a W. M. Keck Foundation grant to the Keck Center for Quantum Magnetism at Rutgers University.
The magnetic susceptibility measurement conducted at NICPB (A.B., J.L., I.H, and R.S.) was supported by the Estonian Science Council Grant PRG1702.
S.M. would like to acknowledge the startup fund from the University of South Carolina. 
This work used the Expanse supercomputer at the San Diego Supercomputer Center through allocation PHY230093 from the Advanced Cyberinfrastructure Coordination Ecosystem: Services \& Support (ACCESS) program, which is supported by National Science Foundation Grants
No.~2138259, No.~2138286, No.~2138307, No.~2137603, and No.~2138296. 

\textbf{Author Contributions} \\
G.B. and T.R. supervised the experiments. 
S.W. and G.B. acquired and analyzed the Raman spectra. 
L.P., U.N. and T.R. acquired and analyzed the terahertz absorption spectra. 
A.B., J.L., I.H., and R.S. acquired and analyzed the magnetic susceptibility data. 
X.X., K.D., and S.C. grew and polished the single crystals. 
Z.W., X.W., and S.M. performed the DFT defect supercell calculations and unfolding analysis. 
All the authors contributed to the discussions and writing of the paper.
    
\textbf{Competing interests} \\
The authors declare no competing interests.

\textbf{Materials \& Correspondence}\\
Correspondence and requests for materials should be addressed to Shangfei~Wu, Sai~Mu, Toomas~Rõõm, and Girsh~Blumberg.

\newpage

%
\renewcommand{\thefigure}{S\arabic{figure}}
\addtocounter{figure}{-4}
\renewcommand{\theequation}{S\arabic{equation}}
\addtocounter{equation}{-0}

\textbf{Supplementary Materials}

\begin{figure}[!b] 
\begin{center}
\includegraphics[width=\columnwidth]{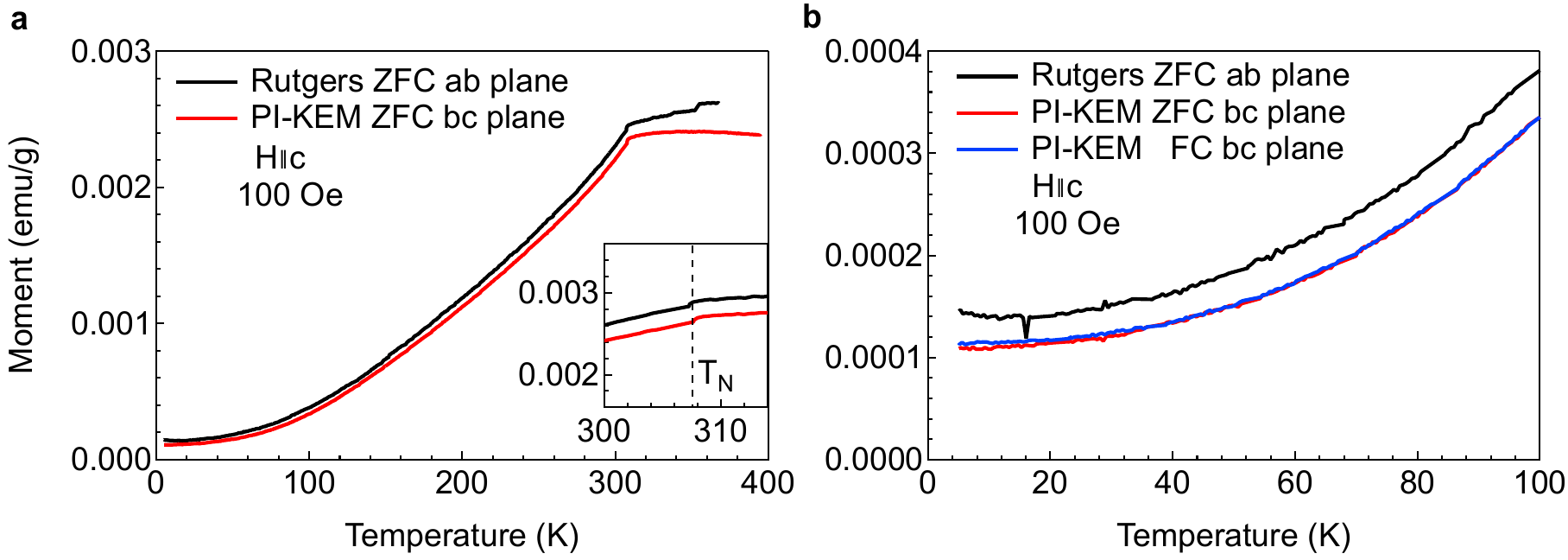}
\end{center}
\caption{\label{Fig.S1} \textbf{Temperature dependence of the magnetic susceptibility data.}
\textbf{a} Magnetization measurements for the ``Rutgers” and ``PI-KEM”-Cr$_2$O$_3$ single crystals with magnetic field along the $c$ axis and $H = 100$\,Oe in the warming-up process. ZFC and FC denote zero-field cooling and field cooling, respectively. The inset of (a) is a zoom of the data around the AFM phase transition. The vertical line represents $T_\text{N}\sim307.5$\,K.
\textbf{b} Same as panel \textbf{a} but for the ZFC and FC measurements below 100\,K.
}
\end{figure}  

\begin{figure}[!b] 
\begin{center}
\includegraphics[width=\columnwidth]{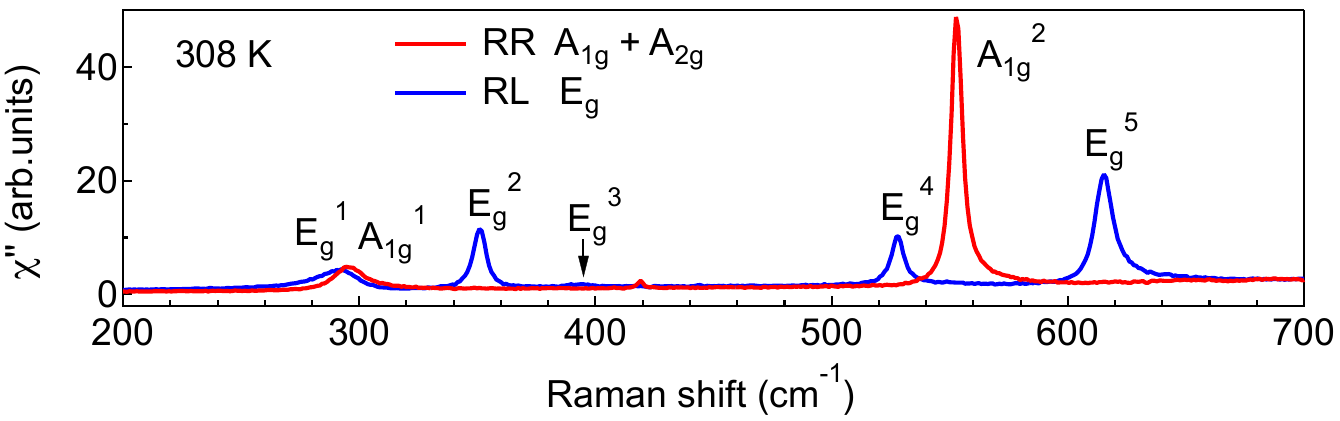}
\end{center}
\caption{\label{Fig.S2} \textbf{Raman response in the $RR$ and $RL$ scattering geometries for the ``Rutgers”-Cr$_2$O$_3$ sample at 308\,K.}
}
\end{figure}

\section{Group-theoretical analysis}\label{GroupTheory}  
                                               
Above $T_N$, Cr$_2$O$_3$ belongs to the hexagonal structure with space group $R\bar{3}c$ (No.~167) [point group $D_{3d}$]. The Cr and O atoms have Wyckoff positions $12c$ and $18e$, respectively.                                                             
From the group theoretical considerations, the $\Gamma$ point phonon modes of the Cr$_2$O$_3$ can be expressed as $\Gamma$ = 2$A_{1g}$ $\oplus$ 2$A_{1u}$ $\oplus$ 3$A_{2g}$ $\oplus$ 3$A_{2u}$ $\oplus$ 5$E_{u}$ $\oplus$ 5$E_{g}$. Raman active modes are $\Gamma_{\text{Raman}}$= 2$A_{1g}$ $\oplus$ 5$E_g$, IR active modes are $\Gamma_{\text{IR}}$ = 2$A_{2u}$ $\oplus$ 4$E_{u}$, and acoustic modes are $\Gamma_{\text{acoustic}}$ = $A_{2u}$ $\oplus$ $E_{u}$. Note that $A_{2g}$ signals are neither Raman nor IR active and might become Raman active under resonant conditions.
For the $D_{3d}$ point group, the Raman selection rules indicate that the $XX$, $XY$, $RR$, and $RL$ polarization geometries probe the $A_{1g} + E_g$, $A_{2g}+ E_g$, $A_{1g} + A_{2g}$, and $2E_g$ symmetry excitations, respectively. The relationship between the scattering geometries and the probed symmetry channels is summarized in Table~\ref{SymmetryAnalysis}. 
The algebra used in this study to decompose the Raman data into three irreducible representations of the point group $D_{3d}$ is summarized in Table~\ref{decompositionD6h}. 
                                            
\begin{table}[b]
\caption{\label{SymmetryAnalysis}The relationship between the scattering geometries and the symmetry channels for point groups $D_{3d}$ and $D_3$. $A_{1g}$,  $A_{2g}$, and $E_{g}$ are the irreducible representations of the $D_{3d}$ point group, while $A_{1}$,  $A_{2}$, and $E$ are the irreducible representations of the $D_{3}$ point group.}
\begin{ruledtabular}
\begin{tabular}{ccc}
Scattering geometry&$D_{3d}$& $D_{3}$\\
\hline
$XX$&$A_{1g}+E_{g}$&$A_{1}+E_{}$\\
$XY$&$A_{2g}+E_{g}$&$A_{2}+E_{}$\\
$RR$&$A_{1g}+A_{2g}$&$A_{1}+A_{2}$\\
$RL$&$2E_{g}$&$2E$\\
\end{tabular}
\end{ruledtabular}
\end{table}

\begin{table}[t]
\caption{\label{decompositionD6h} Algebra used to decompose the Raman data into three irreducible representations of the point group $D_{3d}$ ($D_3$).}
\begin{ruledtabular}
\begin{tabular}{cc}
Symmetry channel $D_{3d}$ ($D_3$)&Expression\\
\hline
$A_{1g}$ ($A_1$)&$\chi''_{XX}-\chi''_{RL}/2$\\
$A_{2g}$ ($A_2$)&$\chi''_{XY}-\chi''_{RL}/2$\\
$E_{g}$ ($E$)&$\chi''_{RL}/2$\\
\end{tabular}
\end{ruledtabular}
\end{table}

Below $T_\text{N}$, the presence of AFM ordering in bulk Cr$_2$O$_3$  breaks $i$, $\sigma_d$, and $S_6$ symmetries associated with the $D_{3d}$ point group symmetries. Consequently, the original $D_{3d}$ point group symmetry is reduced to $D_3$ below $T_\text{N}$, resulting in interirrep mixing between bare phonons of different irreducible representations (irreps). The $E_g$ and $E_u$ irreps become the $E$ irrep, the $A_{1g}$ and $A_{1u}$ become the $A_1$ irrep, and the $A_{2g}$ and $A_{2u}$ irreps become the $A_2$ irrep of the $D_3$ point group [Table~\ref{decompositionD3dD3}].
\textcolor{black}{From the group theoretical considerations, $\Gamma$ point optical phonon modes of the Cr$_2$O$_3$ can be expressed as $\Gamma_{\text{optical}}$ = 4$A_1$ + 6$A_2$ + 10$E$ and the acoustic modes are $\Gamma_{\text{acoustic}}$ =$A_2$ $\oplus$ $E$. 
All the $E$-symmetry optical modes are both Raman and infrared active. The $A_1$-symmetry modes are Raman active, while the $A_2$-symmetry modes  become infrared active.}

\begin{table}[b]
\caption{\label{decompositionD3dD3} The relationship of the irreducible representations between $D_{3d}$ and $D_3$ point groups.}
\begin{ruledtabular}
\begin{tabular}{cc}
$D_{3d}$&$D_3$\\
\hline
$A_{1g}$&$A_1$\\
$A_{1u}$&$A_1$\\
$A_{2g}$&$A_2$\\
$A_{2u}$&$A_2$\\
$E_{g}$&$E$\\
$E_{u}$&$E$\\
\end{tabular}
\end{ruledtabular}
\end{table}

 \section{The symmetry of quasielastic response}\label{QEP} 
\begin{figure}[!t] 
\begin{center}
\includegraphics[width=0.6\columnwidth]{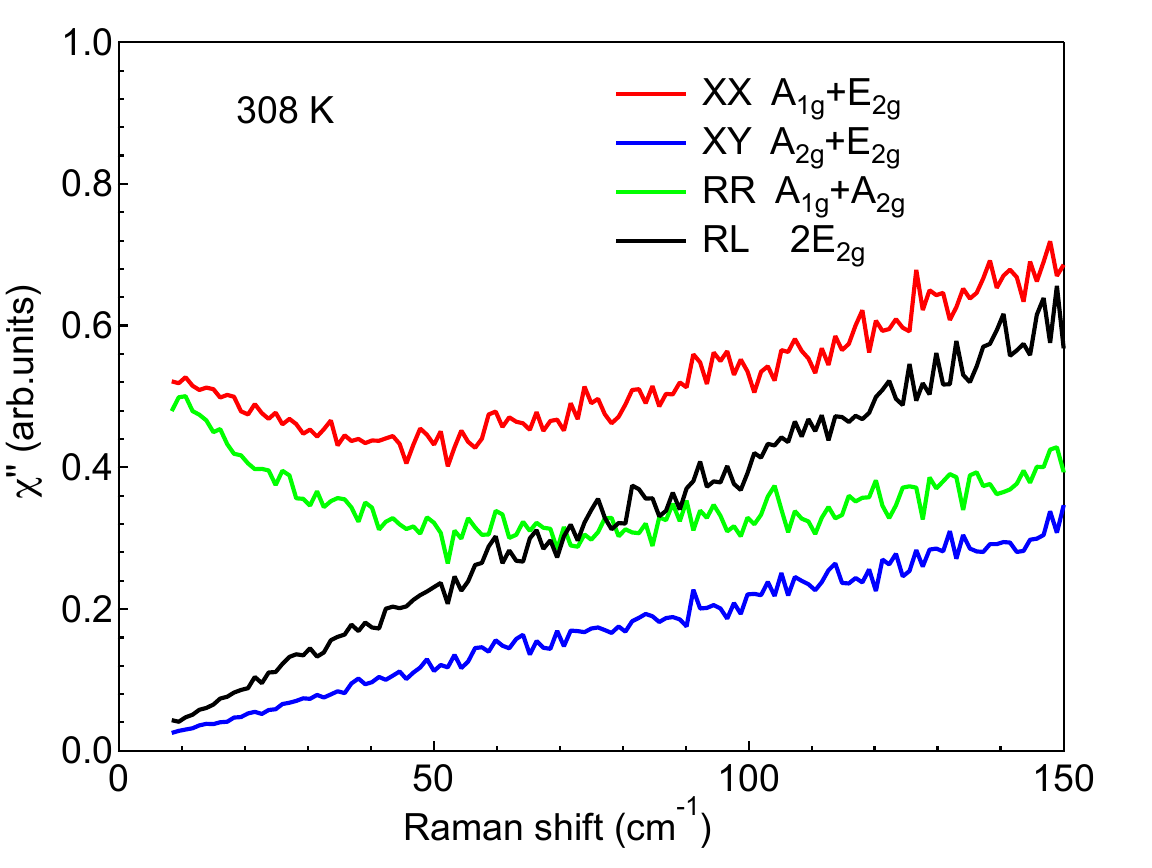}
\end{center}
\caption{\label{Fig.S3}  
\textbf{Quasielastic Raman response.} Polarization dependence of the quasielastic Raman response below 150\,\cm-1 recorded from the $ab$ plane of the ``Rutgers"-Cr$_2$O$_3$ at 308\,K and 0\,T.
 }
\end{figure}

In this section, we show the Raman response of the quasielastic response below 150\,\cm-1 recorded from the $ab$ plane of the “Rutgers”-Cr$_2$O$_3$ at 308\,K and 0\,T in Fig.~\ref{Fig.S3}. The quasielastic scattering intensity appears both in the $XX$ and $RR$, but not in the $XY$ and $RL$ scattering geometries. Based on the selection rule [Table~\ref{SymmetryAnalysis}], the quasielastic scattering signal has the $A_{1g}$ ($A_1$) symmetry according to the $D_{3d}$ ($D_{3}$) point group notation.
     
\section{Fitting of the Fano lineshaped one-magnon mode}\label{fitting_magnon} 
In this section, we show the details of model fitting of the Fano lineshape of the one-magnon mode below 10\,\cm-1.
We consider the interference between a discrete one-magnon mode and an interacting continuum in the same symmetry channel~\cite{Fano1961PhysRev,CardonaBookI,blumberg1994JSC}. 
The bare Raman response of a discrete one-magnon mode is described by a Lorentzian with frequency $\omega_m$, HWHM $\gamma$. It has the form of $ \gamma_m/((\omega_m-\omega)^2+\gamma_m^2)$.
The electronic continuum has the form of $-R(\omega)+ i \rho(\omega)$, where its real and imaginary parts, $R(\omega)$ and $\rho(\omega)$, are connected by the Kramers-Kronig relations.        
Following Klein's approach~\cite{CardonaBookI,blumberg1994JSC}, 
the Raman response of the perturbed system can be obtained as follows:
\begin{equation}
\begin{split}
\chi''(\omega)=&t_e^2 [\pi \rho(\omega)(\omega_0-\omega-vt_{m}/t_e)^2+\gamma(vR(\omega)\\
 &-t_{m}/t_e)^2+\pi \rho(\omega)\gamma(v^2\pi \rho(\omega)+\gamma)]\\
  &/[(\omega_0- \omega + v^2R(\omega))^2 + (v^2 \pi \rho(\omega) +\gamma)^2]
 \end{split}\label{FanoLongVersion}
\end{equation} 
where $v$ is the coupling constant between the one-magnon and the electronic continuum; $t_m$ and $t_e$ are the light coupling amplitudes to the one-magnon and to the electronic continuum in the $E$-symmetry channel, respectively.
To describe the electronic continuum in the Raman data, we find it sufficient to assume a purely relaxational dynamics corresponding to a strong overdamping case, which has a Drude-like formalism: 
     
\begin{equation}
R(\omega) \sim \frac{t_e^2 \gamma_e}{\gamma_e^2+\omega^2}, 
\rho(\omega) \sim \frac{t_e^2 \omega}{\gamma_e^2+\omega^2}
\end{equation} 
where $t_e$ controls the overall intensity determined by the light-scattering vertex. $\gamma_e$ is the relaxation rate.      
 \begin{figure}[!t] 
\begin{center}
\includegraphics[width=0.7\columnwidth]{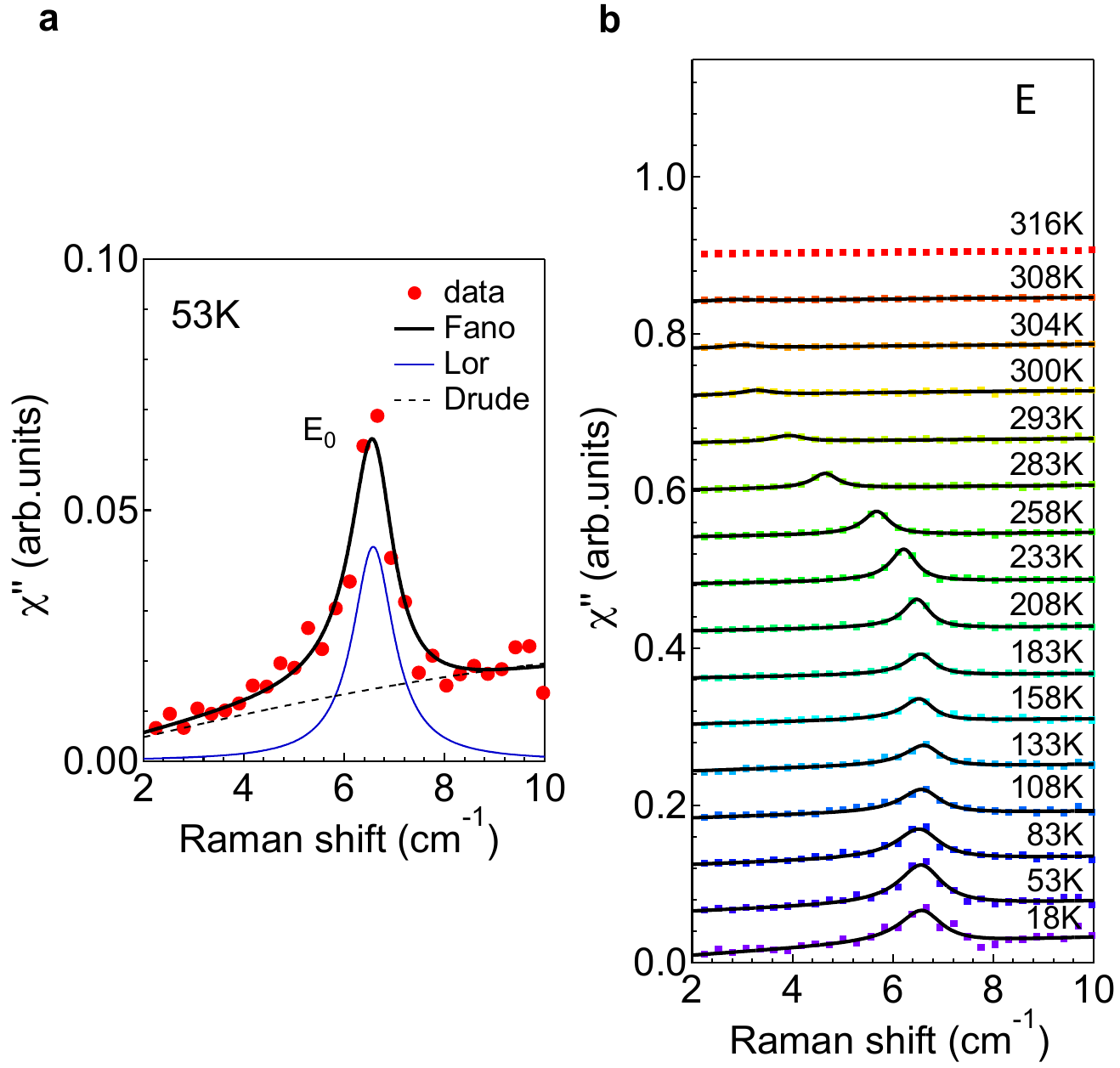}
\end{center}
\caption{\label{Fig.S4}  \textbf{Fitting results.}
\textbf{a} An example of the Fano fitting of the one-magnon peak at 53\,K in the $RL$ scattering geometry. The red dots are the data points. The solid black curve is the coupled Fanoshaped magnon-continuum response. The solid blue and dashed black curves are the bare Lorentzian-like magnon mode and Drude-like electronic continuum response, respectively. \textbf{b} Temperature dependence of the fitted one-magnon curves in the $RL$ scattering geometry based on Eq.~(\ref{FanoLongVersion}).}
\end{figure}   

\begin{figure}[!t] 
\begin{center}
\includegraphics[width=0.6\columnwidth]{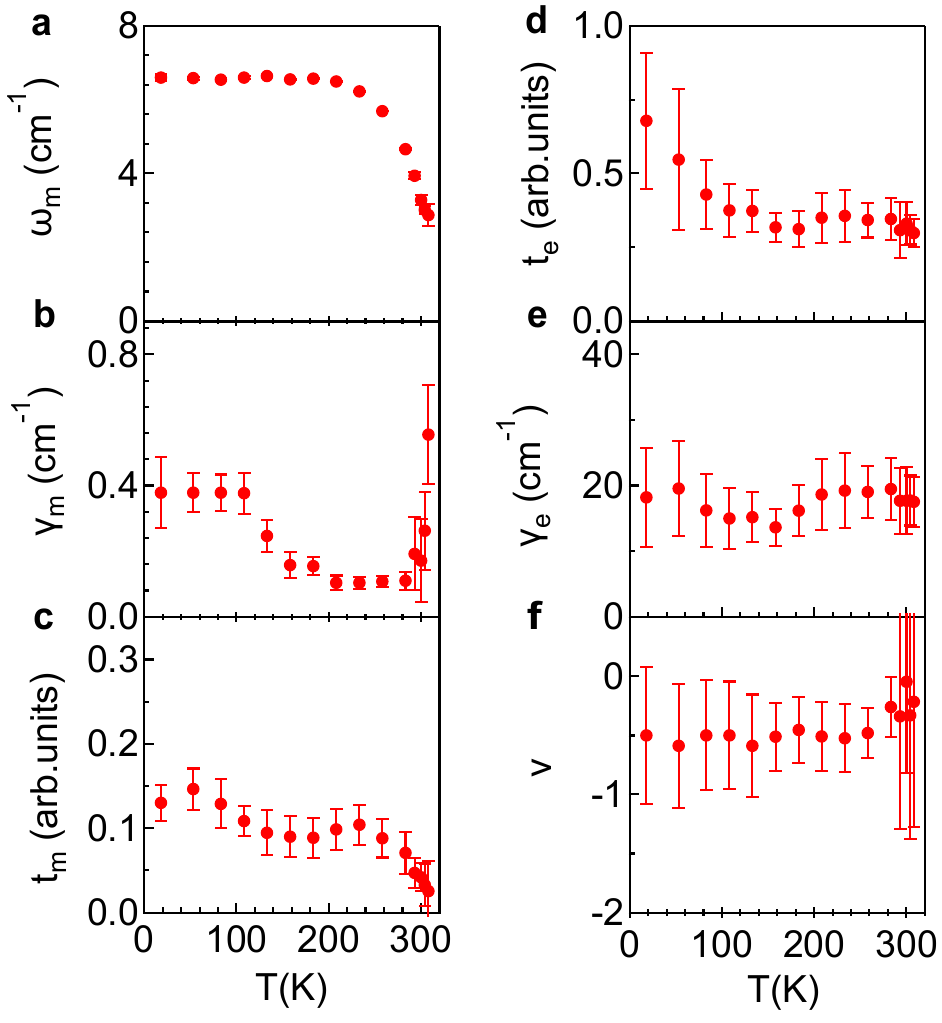}
\end{center}
\caption{\label{Fig.S5}  
\textbf{Temperature dependence of the fitting parameters for the Fano lineshaped one-magnon mode.} 
\textbf{a} One-magnon mode frequency $\omega_m$. \textbf{b} One-magnon mode HWHM $\gamma_m$. \textbf{c} The light coupling anplitude to the one-magnon mode $t_m$. \textbf{d} The light coupling anplitude to the electronic continuum $t_e$. \textbf{e} The relaxation rate for the electronic continuum $\gamma_e$. \textbf{f} The coupling strength $v$.
}
\end{figure} 

We employ the Fano model [Eq.~\ref{FanoLongVersion}] to fit the
asymmetric one-magnon mode. In Fig.~\ref{Fig.S4}a, we show an example of the Fano fitting of the one-magnon scattering peak at 53\,K. The bare  magnon mode, the bare electronic continuum, and the coupled magnon-continuum Fano responses are presented in Fig.~\ref{Fig.S4}a. The model fits the data very well. In Fig.~\ref{Fig.S4}b, we present the temperature dependence of the fitted curves for the one-magnon mode below 10\,\cm-1. The agreement between the data points and the model justifies the Fano model [Eq.~\ref{FanoLongVersion}].

In Fig.~\ref{Fig.S5}, we present the temperature dependence of the fitting parameters for the Fano model Eq.~\ref{FanoLongVersion}. The mode frequency $\omega_m$, HWHM $\gamma_m$, and the light coupling amplitude to the one-magnon mode $t_m$ are shown in Figs.~\ref{Fig.S5}a, \ref{Fig.S5}b, and \ref{Fig.S5}c, respectively. The light coupling amplitude to the electronic continuum $t_e$, and the relaxation rate for the electronic continuum $\gamma_e$, are shown in Figs.~\ref{Fig.S5}d and \ref{Fig.S5}e, respectively. The coupling constant between the one-magnon mode and the electronic continuum $v$ is shown in Fig.~\ref{Fig.S5}f. From these fitting parameters, the one-magnon mode frequency $\omega_m$ behaves like an order parameter below $T_\text{N}$. The HWHM for the magnon $\gamma_m$ first decreases below $T_\text{N}$ but increases substantially below $T^* \sim 150$\,K.
Both the light coupling amplitude to the magnon $t_m$ and to the electronic continuum $t_e$ increase a bit below $T_\text{N}$. The relaxation rate for the electronic continuum $\gamma_e$ and the coupling constant $v$ show little temperature dependence below $T_\text{N}$. 

\section{Threefold rotational symmetry for the defect modes $E^1$, $E^2$, and $E^3$}
\label{three_fold_rotational_symmetry} 

In this section, we show that the threefold rotational symmetry is preserved for the defect modes $E^1$, $E^2$, and $E^3$ in the AFM phase of Cr$_2$O$_3$ by Raman and terahertz absorption spectroscopies.

\begin{figure*}[!t] 
\begin{center}
\includegraphics[width=\columnwidth]{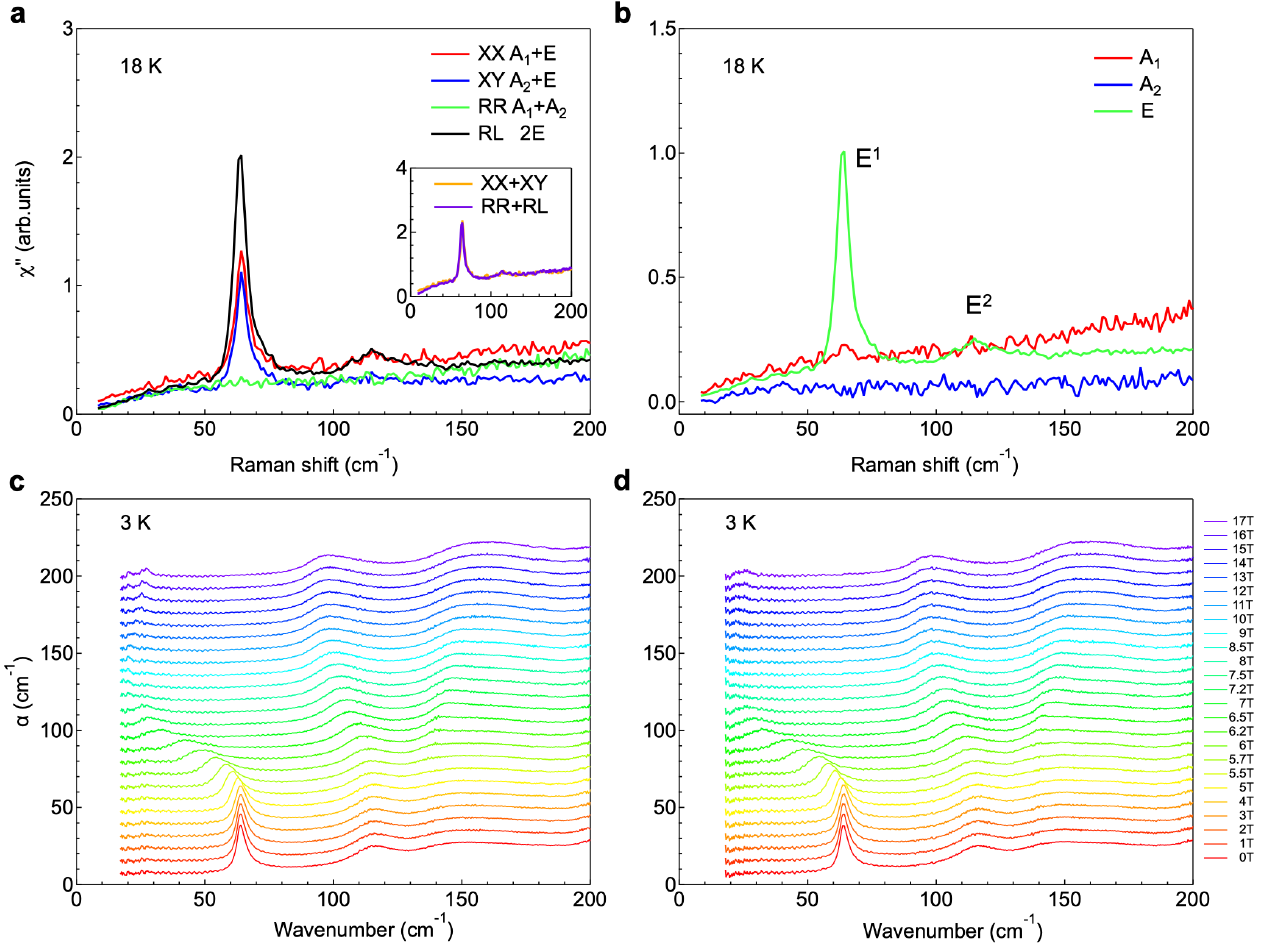}
\end{center}
\caption{\label{Fig.S6}  
\textbf{Threefold rotational symmetry for the defect modes in the AFM phase of ``Rutgers"-Cr$_2$O$_3$.} 
\textbf{a} Raman response in the $XX$, $XY$, $RR$, and $RL$ scattering gemoetries at 18\,K from the $ab$ plane of ``Rutgers"-Cr$_2$O$_3$. The inset of panel \textbf{a} shows the sum of the Raman response in two orthogonal scattering gemoetries $XX+XY$ and $RR+RL$.
\textbf{b} Decompositions of the Raman response into $A_1$, $A_2$, and $E$ irreducible representations of $D_3$ point group using the algebra shown in Table~\ref{decompositionD6h}.
\textbf{c-d} Magnetic field dependence of the terahertz absorption spectra in the Faraday configuration for two orthogonal polarizations in the $ab$ plane. The data shown in Fig.~\ref{Fig3}c of the main text is reploted here in panel \textbf{c}.
}
\end{figure*}

In Fig.~\ref{Fig.S6}a, we present the Raman response in the $XX$, $XY$, $RR$, and $RL$ scattering geometries at 18\,K from the $ab$ plane of “Rutgers”-Cr$_2$O$_3$. In the inset of  Fig.~\ref{Fig.S6}a, we show the sum of the Raman response in two orthogonal scattering geometries $XX + XY$ and $RR + RL$. The nearly identical responses $XX + XY$ and $RR + RL$ justify the internal data consistency. In Fig.~\ref{Fig.S6}b, we show the decomposition of the Raman response into $A_1$, $A_2$, and $E$ irreducible representations of the $D_3$ point group using the algebra given in Table~\ref{decompositionD6h}. The defect modes $E^1$ and $E^2$
have well-defined $E$ symmetry.
Since the selection rule [Table~\ref{SymmetryAnalysis}], sum rule [inset of Fig.~\ref{Fig.S6}a], and decomposition algebra [Table~\ref{decompositionD6h}] are characteristic properties of a lattice system with trigonal or hexagonal symmetry, we conclude that the threefold rotational symmetry is preserved for the defect modes in the AFM phase of Cr$_2$O$_3$. 


In Figs.~\ref{Fig.S6}b,c, we show the terahertz absorption spectra of “Rutgers”-Cr$_2$O$_3$ in the Faraday configuration for two orthogonal polarizations at 3\,K. The data in two orthogonal polarizations are consistent and
similar, ruling out threefold rotational symmetry breaking for the defect modes in the AFM phase of Cr$_2$O$_3$.

\begin{figure}[!t] 
\begin{center}
\includegraphics[width=0.5\columnwidth]{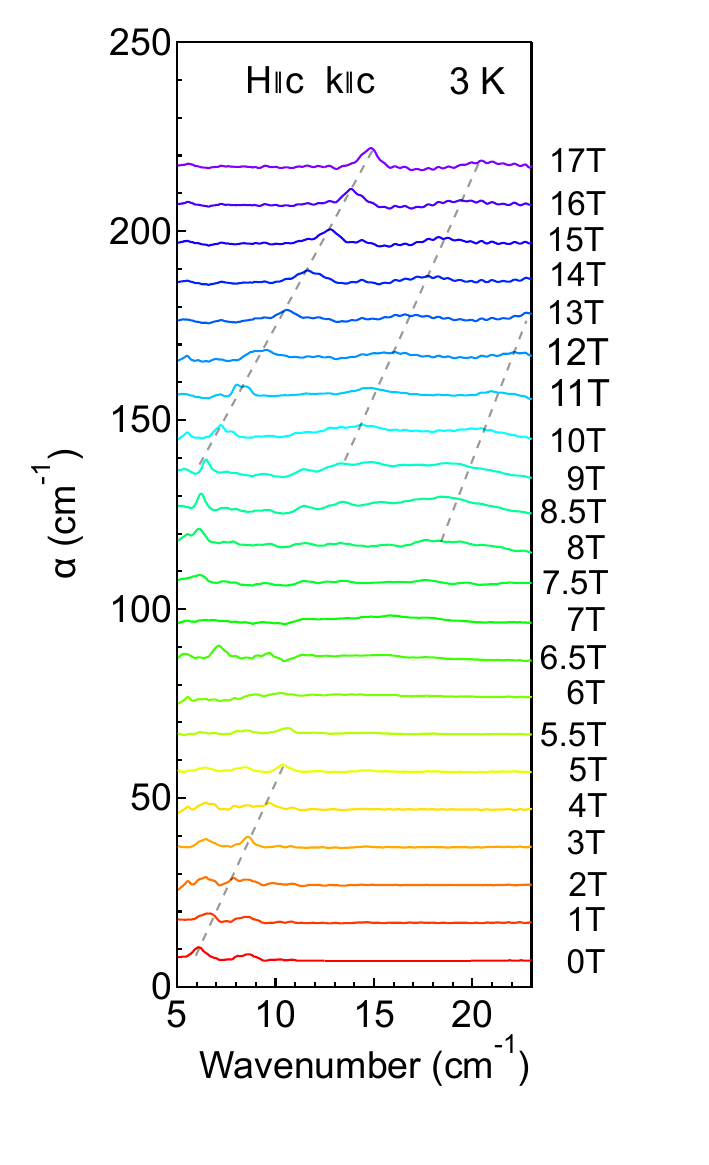}
\end{center}
\caption{\label{Fig.S7}  
Magnetic field dependence of the terahertz  absorption spectra in the Faraday configuration for the ``Rutgers"-Cr$_2$O$_3$ at 3\,K. The dashed lines are guides to the eyes.
}
\end{figure}

\section{Laser heating determination} \label{laser_heating_determination}                                                                                                                             
The laser heating rate, a measure of the temperature increase per unit laser power in the focused laser spot (K/mW), in the Raman experiments was determined by monitoring 
the appearance of one-magnon scattering peak during the laser heating process at room temperature 296\,K. 
With laser power 10\,mW, there is a clear one-magnon scattering peak at  3\,\cm-1, indicating the laser spot temperature is below $T_\text{N}$=307.5\,K.
When heating with laser power of 15\,mW, the one-magnon mode weakens significantly and shifts to lower energy 2.8\,\cm-1, indicating the laser spot temperature is slightly below $T_\text{N}$=307.5\,K.
When heating with laser power of 25\,mW, the one-magnon mode disappears, indicating the laser spot temperature is above $T_\text{N}$=307.5\,K. Thus, the heating coefficient can be determined via the constraint: $296\,\text{K}+15\,\text{mW}*k  \approx 307.5$\,K. In this way, we have deduced the heating coefficient: $k \approx 0.77\pm0.1$\,K/mW.

\begin{figure}[!t] 
\begin{center}
\includegraphics[width=1\columnwidth]{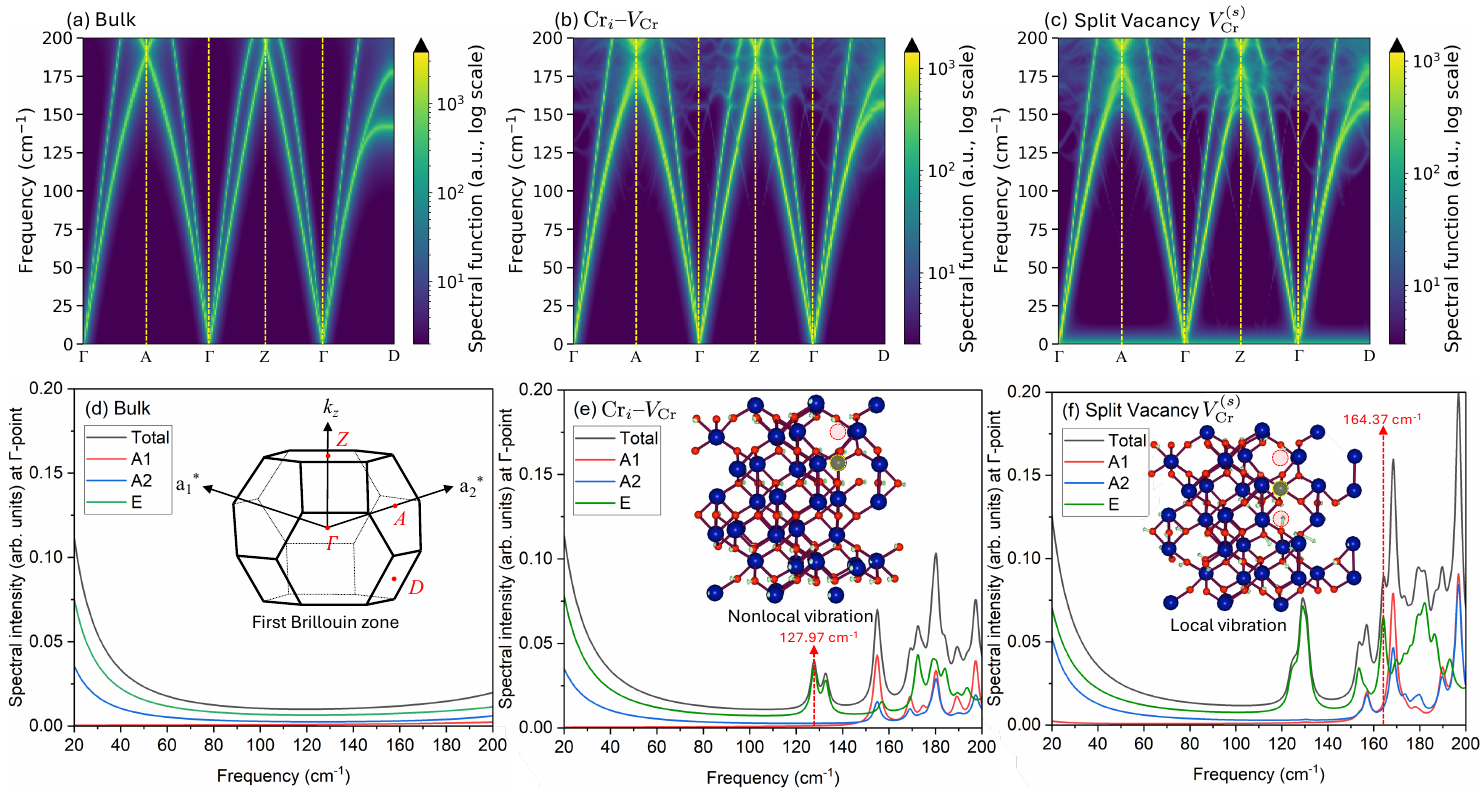}
\end{center}
\caption{\label{Fig.S8}  
\textbf{DFT defect supercell calculations.} Phonon spectral functions (arbitrary units) of bulk Cr$_2$O$_3$ (panel \textbf{a}), Cr$_i$-$V_\text{Cr}$ (panel \textbf{b}), and $V^{(s)}_\text{Cr}$ (panel \textbf{c}), obtained from supercell phonon unfolding method. The spectral functions are given up to 200\,cm$^{-1}$ and high-symmetry $k$ points are labeled in the first Brillouin Zone [inset of  panel \textbf{d}]. Spectral intensities at the $\Gamma$ point and their decompositions into $A_1$, $A_2$ and $E$-symmetry channels as a function of frequency are shown for bulk Cr$_2$O$_3$ (panel \textbf{d}), Cr$_i$-$V_\text{Cr}$ (panel \textbf{e}), and $V^{(s)}_\text{Cr}$ (panel \textbf{f}). Inset of panels \textbf{e} and \textbf{f} illustrate the eigenmodes of vibrations at 127.97 cm$^{-1}$ and 164.37 cm$^{-1}$ therein, respectively. The interstitial Cr atoms and the Cr vacancy sites are marked with grey and pink balls [inset of panels \textbf{e}-\textbf{f}], respectively.
}
\end{figure}

\section{Calculation of the local magnetic excitation energy at the ground state position} \label{local_spin_flip_enegy}   

\textcolor{black}{  
We build up a spin Hamiltonian for the interstitial Cr within the split vacancy $V^{(s)}_\text{Cr}$:
\begin{equation}
H= -\sum_i J_i \boldsymbol{S} \cdot \boldsymbol{s}_i-D S_z^2
\end{equation}
Here, $\boldsymbol{S}$ is the interstitial Cr spin (with $\boldsymbol{S} = 3/2$), $S_z$ is its projection along the quantization axis, $s_i$ represents the nearby Cr spins, and $J_i$ is the Heisenberg exchange interaction between the interstitial Cr and the neighboring Cr atom $i$. $D$ is the magnetic anisotropy parameter. The threefold rotational symmetry is preserved in the presence of interstitial Cr, so only the $-D S_z^2$ term survives constrained by symmetry. }
 
\begin{figure}[!t] 
\begin{center}
\includegraphics[width=1\columnwidth]{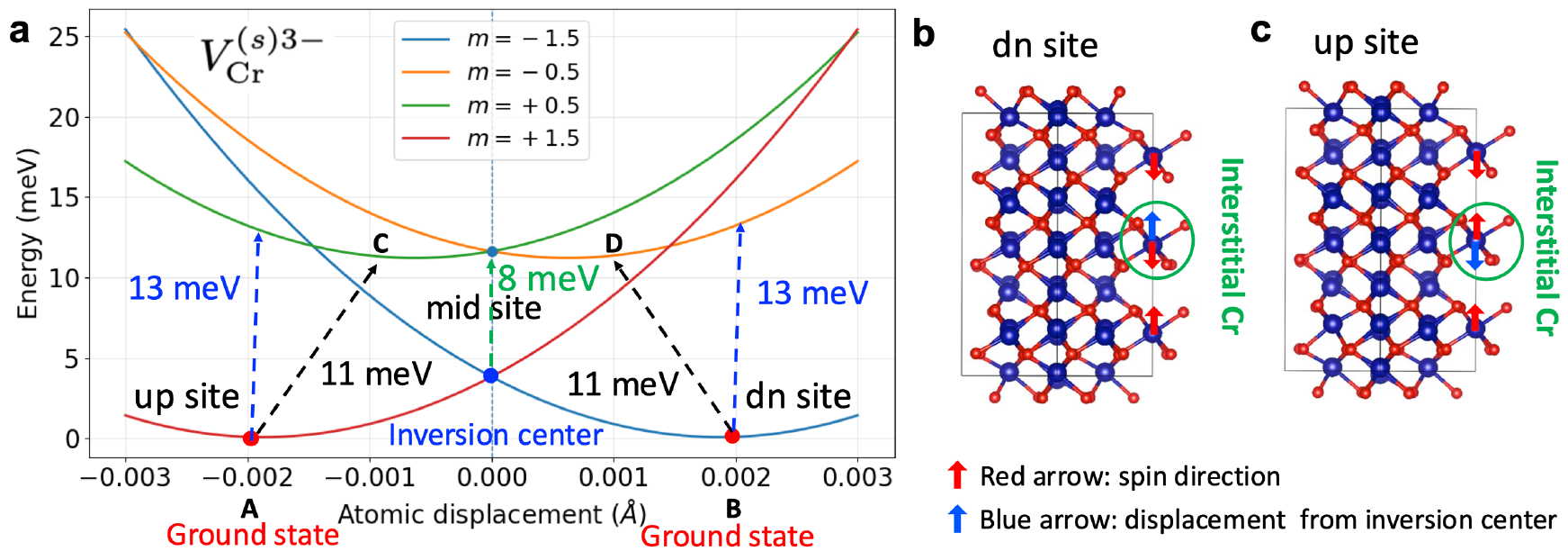}
\end{center}
\caption{\label{Fig.S9}  
\textbf{Calculation of the local magnetic excitation energy at the ground state position for the split-vacancy $V^{(s)3-}_\text{Cr}$.} 
\textbf{a} Energy diagram of $V^{(s)3-}_{\text{Cr}}$ as a function of the interstitial Cr position, between the degenerated spin-up and spin-down ground states.  The two minima positions of the $S_z = \pm 3/2$  parabolas are labled as A and B, while the two minima positions of the $S_z = \pm 1/2$  parabolas are labed as C and D.
\textbf{b} The split-vacancy $V^{(s)}_\text{Cr}$ with a spin-down and upward displacement for the interstitial Cr atom.
\textbf{c} The split-vacancy $V^{(s)}_\text{Cr}$ with a spin-up and downward displacement for the interstitial Cr atom. 
}
\end{figure}

 \textcolor{black}{     
The effective exchange field acting on the interstitial Cr can be defined as $\sum_i J_i s_i = J_0 (\bold{R_0})$, which depends on the position of the interstitial Cr at position $\bold{R_0}$ ($\bold{R_0}$ is the distance between the interstitial Cr and the inversion center along the easy-axis direction within the split Cr vacancy $V^{(s)}_\text{Cr}$). In the mean-field limit, energies for different $S_z$ from the Hamiltonian are:  
\begin{equation}
\label{eq:equation}
\begin{split}
E(S_z = \pm 3/2) &= \pm \frac{3}{2} J_0(\bold{R_0})-\frac{9}{4}D,\\
E(S_z = \pm 1/2) &= \pm \frac{1}{2} J_0(\bold{R_0})-\frac{1}{4}D.
\end{split}
\end{equation}
$J_0$ can be estimated from the energy difference required to flip the direction of the
 interstitial Cr spin with the structure frozen and the spins on the rest Cr sites frozen. 
We compute $J_0$ for the two equivalent ground states (minimum of the two $S_z = \pm 3/2$ parabolas shown in Fig.~\ref{Fig.S9}a) and get  $J_0 = \pm 5.3$\,meV for the two minima. We then linearly interpolate $J_0$ for the interstitial Cr position between the two ground states. }

\textcolor{black}{  
For the interstitial Cr located at the inversion center ($\bold{R_0}=0$), $J_0=0$\,meV constrained by symmetry, the energy difference between the $S_z = \pm 3/2$ and $S_z = \pm 1/2$ states is therefore $2D$ (see Eq.~\ref{eq:equation}).
For the interstitial Cr located at the ground state position, the energy difference between the $S_z = \pm 3/2$ and $S_z = \pm  1/2$ states is  $|J_0|+2D$  (see Eq.~\ref{eq:equation}).
}

\textcolor{black}{  
To estimate the anisotropy $D$ parameter, we calculate the variation in full relativistic energy for the defect supercell when the interstitial Cr spin direction is changed from the easy axis direction to in-plane direction. 
The calculated single-ion anisotropy energy $(3/2)^2 D$ is 9\,meV, we then deduce that $D$ is 4\,meV.  }

\textcolor{black}{  
Therefore, $2D=8$\,meV corresponds to the magnetic excitation energy for the interstitial Cr spin excited by light, if the interstitial Cr spin is located precisely at the inversion center. 
To estimate the magnetic excitation energy at the ground state position, we assume that $D$ is constant near the inversion center,  and we further incorporate the elastic interactions from the fitted quadratic terms based on the parabola shown in Fig.~5 of Ref.~\cite{Sai_Mu_2025_arxiv}. When combined with the interpolated $J_0$, the energy profile as a function of interstitial Cr positions for all $S_z$ is illustrated in Fig.~\ref{Fig.S9}. The magnetic excitation energy for the interstitial Cr spin excited by light, corresponding to the energy difference between the two minima of the $S_z = + 3/2$  and $S_z = + 1/2$ parabolas (optical transition from A to C) shown in Fig.~\ref{Fig.S9}a, or equivalently between the two minima of the $S_z =- 3/2$  and $S_z = -1/2$ parabolas (optical transition from B to D), is about 11\,meV. 
}

\newpage
\section{Selection rule of the local magnetic excitation and the field independence for low magnetic fields} \label{tunneling}  

\textcolor{black}{
As shown in Fig.~\ref{Fig.S9}a, the distance $d$ between two ground positions (labeled as A and B) along the $c$ axis is about 0.004~\AA, which is two orders of magnitude smaller than the Bohr radius. By applying the WKB (Wentzel–Kramers–Brillouin) approximation, one can estimate the wavevector $k$ inside the potential barrier using the relation $k \sim \sqrt{2 m V}/\hbar$ in a quantum tunneling problem, where $V$ is the potential barrier, $m$ is the mass, and $\hbar$ is the reduced Planck constant.
For a classical Cr atom with 52 atomic mass units and a few millielectronvolt potential barrier deduced from Fig.~\ref{Fig.S9}a, the wavevector $k$ inside the potential barrier is about 5 \AA$^{-1}$. Then, the tunneling exponent $\gamma \sim k\cdot d$ is about 5~\AA$^{-1} \times 0.004$~\AA = 0.02. Thus, the transmission coefficient for an interstitial Cr atom tunnelling through the potential barrier located at the inversion center from positions A to B (Fig.~\ref{Fig.S9}a) calculated via $e^{-2\gamma}$ is found to be close to 1.
The consequence of this spin-flip tunneling is that (1) the very close parabolas with opposite spin directions almost merge into a single potential well; (2) the spin direction of the interstitial Cr atom within the  split vacancy $V^{(s)}_\text{Cr}$ fluctuates very rapidly.
}
 
\subsection{Tunneling Hamiltonian}

To describe a single spin-3/2 interstitial Cr tunneling between two ground-state positions (labeled as A and B) with the same energy but opposite spin directions, we use a composite Hilbert space that multiplies its spatial position by its internal spin degrees of freedom. This system forms a Hilbert space governed by spin-position entanglement.

We first define the spin projections and the state vectors.
For a spin-3/2 interstitial Cr, the magnetic quantum number $m$ has four possible values: $+3/2, +1/2, -1/2, -3/2$. The ``opposite spin direction" means the spin projection is exactly reversed ($m \rightarrow -m$). This pairs the states into two distinct tunneling channels:
\begin{equation}
\begin{split}
\text{Channel 1}: \vert +3/2 \rangle \longleftrightarrow \vert -3/2 \rangle\\
\text{Channel 2}: \vert +1/2 \rangle \longleftrightarrow \vert -1/2 \rangle
\end{split}
\end{equation}
We define the two ground-state positions $A$ and $B$ for Channel 1, and define the two positions $C$ and $D$ for Channel 2 shown in Fig.~\ref{Fig.S9}a.
Depending on which spin state the interstitial Cr starts with, the coherent quantum state $\vert \Psi(t) \rangle$ evolves within one of two independent subspaces:

Case 1: The $\pm 3/2$ subspace

The interstitial Cr tunnels between position $A$ with spin $+3/2$ and position $B$ with spin $-3/2$. The state is a superposition:
\begin{equation}
|\Psi _{3/2}\rangle =c_{A}|A,+3/2\rangle +c_{B}|B,-3/2\rangle. 
\end{equation}

Case 2: The $\pm 1/2$ subspace

The interstitial Cr tunnels between position $C$ with spin $+1/2$ and position $D$ with spin $-1/2$. The state is a superposition:
\begin{equation}
|\Psi _{1/2}\rangle =d_{C}|C,+1/2\rangle +d_{D}|D,-1/2\rangle.
\end{equation}
(Note: $c_A$, $c_B$, $d_C$, and $d_D$ are complex probability amplitudes where the sum of their absolute squares for each pair equals 1, such as $c_A^2+c_B^2=1$, $d_C^2+d_D^2=1$).

We then discuss the tunneling Hamiltonian matrix.
The interstitial Cr at positions A and B share the same energies. It is the same for positions C and D. We define that the interstitial Cr has a base energy of \(E_{0}\)  at positions A and B with spin \(\pm 3/2\), and  \(E_{1}\) at positions C and D with spin  \(\pm 1/2\).
 Let \(\Delta _{3/2}\) be the tunneling amplitude between the \(\pm 3/2\) states, and \(\Delta _{1/2}\) be the tunneling amplitude between the \(\pm 1/2\) states.

In the basis \(\{\vert A, +3/2 \rangle, \vert B, -3/2 \rangle, \vert C, +1/2 \rangle, \vert D, -1/2 \rangle\}\), the Hamiltonian blocks out into two independent \(2 \times 2\) matrices:
\begin{equation}
\begin{split}\label{Hamiltonian1}
H_0=\left(
\begin{matrix}
E_{0}&\Delta _{3/2}&0&0\\ 
\Delta _{3/2}&E_{0}&0&0\\ 
0&0&E_{1}&\Delta _{1/2}\\
0&0&\Delta _{1/2}&E_{1}
\end{matrix}
\right)
\end{split}
\end{equation}

By diagonalizing this Hamiltonian (Eq.~\ref{Hamiltonian1}), the eigenvectors of this Hamiltonian give us four stable states. These are maximally spin-position entangled states:

For the \(\pm 3/2\) Channel:
\begin{equation}
|\Psi _{3/2}^{\pm }\rangle =\frac{1}{\sqrt{2}}\left(|A,+3/2\rangle \pm |B,-3/2\rangle \right)\quad \text{with\ energies\ }E_{3/2}^{\pm}=E_{0}\pm \Delta _{3/2}
\end{equation}

For the \(\pm 1/2\) Channel:
\begin{equation}
|\Psi _{1/2}^{\pm }\rangle =\frac{1}{\sqrt{2}}\left(|C,+1/2\rangle \pm |D,-1/2\rangle \right)\quad \text{with\ energies\ }E_{1/2}^{\pm}=E_{1}\pm \Delta _{1/2}
\end{equation}

If \(\Delta_{1/2} >0\) and \(\Delta_{3/2} > 0\), $|\Psi _{3/2}^{+}\rangle$  and $|\Psi _{1/2}^{+}\rangle$
are the symmetric states with higher energies, while $|\Psi _{3/2}^{-}\rangle$  and $|\Psi _{1/2}^{-}\rangle$ are the antisymmetric states with lower energies. 

Based on Fig.~\ref{Fig.S9}a, $|\Psi _{3/2}^{-}\rangle$ is the ground state considering the spin-flip tunneling process for the interstitial Cr.

\subsection{Selection rule of the excitation}

Since the distance between two ground positions (labeled as A and B) along the c axis is about 0.004~\AA~and the potential barrier deduced from Fig.~\ref{Fig.S9}a is a few millielectronvolts, we may treat the spatial and spin parts of the wave-function for the interstitial Cr independently. The spin parts of the wave-function define the selection rules for the local magnetic excitation for the interstitial Cr.

For optical absorption, the selection rule is $\Delta \hat{S}_z = \pm 1$ for electric-dipole active transitions. The corresponding dipole operator $\hat{IR}$ for matrix elements is proportional to $\hat{S}_+$ or $\hat{S}_-$ ($\hat{S}_\pm=\hat{S}_x\pm i \hat{S}_y$).

Following this rule, we obtain:
\begin{equation}\label{IR}
\begin{split}
\langle \Psi _{3/2}^{-}| \hat{IR} |\Psi _{1/2}^{-} \rangle &=1/2 \cdot (\langle A, +3/2| \hat{IR} |C,+1/2\rangle+\langle B, -3/2| \hat{IR} |D,-1/2\rangle)\\
\langle \Psi _{3/2}^{-}| \hat{IR} |\Psi _{1/2}^{+} \rangle &=1/2 \cdot (\langle A, +3/2| \hat{IR} |C,+1/2\rangle-\langle B, -3/2| \hat{IR} |D,-1/2\rangle)\\
\langle \Psi _{3/2}^{-}| \hat{IR} |\Psi _{3/2}^{+} \rangle &=0\\
\langle \Psi _{1/2}^{-}| \hat{IR} |\Psi _{1/2}^{+} \rangle &=0 \\
\langle \Psi _{1/2}^{-}| \hat{IR} |\Psi _{3/2}^{+} \rangle &= 1/2 \cdot(\langle A, +3/2| \hat{IR} |C,+1/2\rangle-\langle B, -3/2| \hat{IR} |D,-1/2\rangle)\\
\langle \Psi _{1/2}^{+}| \hat{IR} |\Psi _{3/2}^{+} \rangle &= 1/2 \cdot(\langle A, +3/2| \hat{IR} |C,+1/2\rangle+\langle B, -3/2| \hat{IR} |D,-1/2\rangle)\\
\end{split}
\end{equation}

\textcolor{black}{For Raman scattering, because light only couples to the electron’s orbital degree of freedom, the effective Raman operator should be written in terms of the orbital angular momentum operators. Spin only becomes visible through a multi-step perturbation process mediated by spin-orbit coupling (SOC).
The incident photon excites an orbital transition (governed by \(\mathbf{\hat{L}}\)). The spin-orbit coupling term (\(\lambda \mathbf{\hat{L}}\cdot\mathbf{\hat{S}}\)) mixes the orbital states with the spin states. The scattered photon is emitted via another orbital transition (governed by \(\mathbf{\hat{L}}\)). Through this indirect mechanism, an effective scattering operator can be mapped onto spin operators (\(\mathbf{\hat{S}}\)) for the ground state.  An effective quadrupole operator $\hat{R}$ for matrix elements after integrating out the orbital and SOC steps is proportional to $\hat{S}_+^2$ or $\hat{S}_-^2$ in terms of pure spin \(\mathbf{\hat{S}}\). The corresponding selection rule is $\Delta \hat{S}_z = \pm 2$ for electric-quadrupole-active transitions, which can be reached in $XY$, $RL$, and $LR$ scattering geometries.}

Following this rule, we obtain:
\begin{equation}\label{Raman}
\begin{split}
\langle \Psi _{3/2}^{-}| \hat{R} |\Psi _{1/2}^{-} \rangle &=1/2\cdot(-\langle A, +3/2| \hat{R} |D,-1/2\rangle-\langle B, -3/2| \hat{R} |C,+1/2\rangle)\\
\langle \Psi _{3/2}^{-}| \hat{R} |\Psi _{1/2}^{+} \rangle &=1/2\cdot(\langle A, +3/2| \hat{R} |D,-1/2\rangle-\langle B, -3/2| \hat{R} |C,+1/2\rangle)\\
\langle \Psi _{3/2}^{-}| \hat{R} |\Psi _{3/2}^{+} \rangle &=0\\
\langle \Psi _{1/2}^{-}| \hat{R} |\Psi _{1/2}^{+} \rangle &=0 \\
\langle \Psi _{1/2}^{-}| \hat{R} |\Psi _{3/2}^{+} \rangle &= 1/2\cdot(-\langle A, +3/2| \hat{R} |D,-1/2\rangle+\langle B, -3/2| \hat{R} |C,+1/2\rangle)\\
\langle \Psi _{1/2}^{+}| \hat{R} |\Psi _{3/2}^{+} \rangle &= 1/2\cdot(\langle A, +3/2| \hat{R} |D,-1/2\rangle+\langle B, -3/2| \hat{R} |C,+1/2\rangle)\\
\end{split}
\end{equation}
              
\textcolor{black}{
Based on Eq.~\ref{IR} and Eq.~\ref{Raman}, the transitions from the ground state $\Psi _{3/2}^{-}$ to the excited state  $\Psi _{1/2}^{-}$ states are both IR and Raman active. This explains why we detect the $E^1$ mode both in the terahertz absorption and Raman spectroscopies.
}

\subsection{Magnetic field dependence}
When an external magnetic field \(\mathbf{B}\) is along the quantization axis (\(c\)-axis), it interacts with the interstitial Cr's magnetic moment via the Zeeman effect. This adds a term to the Hamiltonian that shifts the energy levels based on the magnetic quantum number \(m\). The interaction Hamiltonian for the external magnetic field is:
\begin{equation}
H_{Z}=-\mathbf{\mu }\cdot \mathbf{B}=g\mu _{B}B_{z}S_{z},
\end{equation}
where \(g\) is the landé \(g\)-factor of the interstitial Cr, \(\mu _{B}\) is the Bohr magneton, \(B_{z}\) is the strength of the magnetic field along the \(c\)-axis, \(S_{z}\) is the spin operator with eigenvalues \(m\hbar\) where \(m \in \{+\tfrac{3}{2}, +\tfrac{1}{2}, -\tfrac{1}{2}, -\tfrac{3}{2}\}\). 

To simplify the notation, we define the characteristic Zeeman energy scaling factor as \(\omega_Z = g \mu_B B_z \hbar\). The energy shift for a given state is simply \(m \omega_Z\).
The Hamiltonian (Eq.~\ref{Hamiltonian1}) is modified to \(H = H_0 + H_Z\)  (Eq.~\ref{Hamiltonian2}): 
\begin{equation}
\begin{split}\label{Hamiltonian2}
H=\left(\begin{matrix}
E_{0}+\frac{3}{2}\omega _{Z}&\Delta _{3/2}&0&0\\
 \Delta _{3/2}&E_{0}-\frac{3}{2}\omega _{Z}&0&0\\
  0&0&E_{0}+\frac{1}{2}\omega _{Z}&\Delta _{1/2}\\
  0&0&\Delta _{1/2}&E_{0}-\frac{1}{2}\omega _{Z}
  \end{matrix}\right)
\end{split}
\end{equation}

Diagonalizing each block independently yields four completely distinct energy levels.

For the \(\pm 3/2\) Channel:
\begin{equation}
E_{3/2}^{\pm }=E_{0}\pm \Delta _{3/2} \sqrt{\left(\frac{3\omega _{Z}}{2\Delta _{3/2}}\right)^{2}+1}
\end{equation}

For the \(\pm 1/2\) Channel:
\begin{equation}
E_{1/2}^{\pm }=E_{1}\pm \Delta _{1/2} \sqrt{\left(\frac{\omega _{Z}}{2\Delta _{1/2}}\right)^{2}+1}
\end{equation}

When \(B_z = 0\) (\(\omega_Z = 0\)), the system has four levels at \(E_0 \pm \Delta_{3/2}\) and \(E_1 \pm \Delta_{1/2}\).

When \(B_z \neq 0\), the magnetic field shifts those energy levels. 

In the limit of  $\Delta_{3/2} >> \omega_Z$ and $\Delta_{1/2} >> \omega_Z$, the magnetic field \(B_z\) (\(\omega_Z\)) plays a marginal role in affecting the energy difference between the \(\pm 3/2\) and \(\pm 1/2\) spin-position entangled states.  They are separated in energy by a gap proportional to \(\vert E_0 - E_1 \vert\). 
This may explain the weak field independence of the $E^1$ mode for low magnetic fields below 5\,T shown in Fig.~3 of the main text.

\end{document}